\documentclass[aps,twocolumn,letterpaper, superscriptaddress]{revtex4-1}

\usepackage{amsmath}
\usepackage{amssymb} 
\usepackage{amsfonts}
\usepackage{graphicx}
\usepackage{dcolumn}
\usepackage{float}
\usepackage{bm}
\usepackage{mathrsfs}
\usepackage{tabularx}
\usepackage{bigstrut}
\usepackage{epstopdf}
\usepackage{xfrac}
\usepackage{listings}
\usepackage[usenames,dvipsnames]{xcolor}
\usepackage[hyperindex, breaklinks,colorlinks = true,linkcolor = blue,urlcolor=blue,citecolor=blue]{hyperref}

\begin{document}

\title{Quantum Interference Control of Carriers and Currents in Zincblende Semiconductors based on Nonlinear Absorption Processes}

\author{Rodrigo A. Muniz}
\altaffiliation[Current address: ]{Department of Electrical Engineering and Computer Science, University of Michigan, Ann Arbor, MI 48109, USA}
\affiliation{Department of Physics, University of Toronto, Toronto, Ontario M5S 1A7, Canada}

\author{Cuauht\'{e}moc Salazar}
\altaffiliation[Current address: ]{Department of Physics, University of Warwick, Coventry, CV4 7AL, United Kingdom}
\affiliation{Department of Physics, University of Toronto, Toronto, Ontario M5S 1A7, Canada}

\author{Kai Wang}
\altaffiliation[Current address: ]{Department of Physics, Texas A\&M University, College Station, TX 77843, USA}
\affiliation{Department of Physics, University of Michigan, Ann Arbor, Michigan 48109, USA}

\author{S. T. Cundiff}
\affiliation{Department of Physics, University of Michigan, Ann Arbor, Michigan 48109, USA}

\author{J. E. Sipe}
\email[email: ]{sipe@physics.utoronto.ca}
\affiliation{Department of Physics, University of Toronto, Toronto, Ontario M5S 1A7, Canada}

\date{\today }

\begin{abstract}
Quantum interference between optical absorption processes can excite
carriers with a polarized distribution in the Brillouin zone depending
on properties of the incident optical fields. The polarized distribution
of carriers introduces a current that can be controlled by the phases
and polarizations of the incident optical fields. Here we study the
quantum interference of 2- and 3-photon absorption processes in AlGaAs.
We present theoretical predictions for carrier and current injection
rates considering different frequencies, phases, and polarizations of the incident fields. 
We also discuss the important features that result from only nonlinear optical processes being involved, which leads for instance to a sharper distribution of carriers in the Brillouin zone. 
\end{abstract}

\maketitle

\section{Introduction }

\label{sec:intro}

Quantum interference between different optical processes arises when two optical beams of different frequencies can lead to the same transition.
It has been used to study photoionization of molecular systems \cite{Manykin,Shapiro03, Brumer86, Chen90,Zhu95, Nagai06}, and asymmetric photoejection in semiconductors \cite{Kurizki89, Baranova90,Lawandy90, Baranova91,Dupont95}.
In a crystal, amplitudes for different optical processes leading to electron-hole excitations can interfere constructively in some regions
of the Brillouin zone (BZ), and destructively in others. 
By controlling the polarizations and phases of the incident fields, it is possible to excite carriers in selected localized regions of the BZ. 
Such Quantum Interference Control (QuIC), using 1- and 2-photon absorption processes (1+2 QuIC), has been used for current injection in semiconductors \cite{Atanasov96,Hache97,Rioux12},
graphene \cite{Sun10,Rioux11,Rao12}, topological insulators \cite{Muniz14,Bas15,Bas16},
and transition metal dichalcogenides \cite{Muniz15,Cui15, Mahon18}, as well
as spin current injection in semiconductors \cite{Bhat00,Stevens02,Stevens03, Hubner03,Zhao06}. 
It has also been theoretically investigated for current injection
in graphene nanoribbons \cite{Salazar16}, spin currents in topological
insulators \cite{Muniz14}, and spin and valley currents \cite{Behnia12} in transition
metal dichalcogenides \cite{Muniz15,Mahon18}. 

In crystalline materials, every instance of QuIC studied to date has involved 1- and 2-photon absorption processes, partly because phase related optical fields of frequencies $\omega$ and $2\omega$ can be conveniently achieved by second harmonic generation, while fractional ratios of the frequencies are harder to obtain. 
Currents injected via 1+2 QuIC have been exploited to determine parameters of the optical fields responsible for their injection \cite{Roos05optlet,Roos05laser,Smith07}.
This method has found application in the measurement \cite{Roos04,Fortier04} and stabilization of the carrier-envelope phase of a train of octave-spanning
laser pulses \cite{Cundiff03,Ye05,Roos05josaB}. 
However, phase-coherent frequency combs can also be used to study more general $M+N$ QuIC, especially for fractional ratios $N/M <2$ that only require a narrower frequency range of the comb.
Thus the use of optical frequency combs for QuIC experiments presents  an opportunity to separately study several nonlinear optical processes in semiconductors, which cannot easily be done using simple harmonic generation, as it only produces frequencies that are integer multiples.

In this article we present a theoretical study of QuIC with 2- and 3-photon processes in AlGaAs. 
The injection rates of 1+2 QuIC correspond to divergences in the third-order electric susceptibility, $\chi^{(3)}$, while the injection rates of 2+3 QuIC correspond to divergences in the fifth-order electric susceptibility, $\chi^{(5)}$.
Although sophisticated calculations of $\chi^{(3)}$ have been performed for similar analyses in 1+2 QuIC \cite{Nastos07,Sternemann13,Sternemann16}, here we use a Fermi Golden Rule approach in the perturbative regime, as it allows us to focus only on the resonant processes we are interested, and ignore many other non-resonant processes described by $\chi^{(5)}$. 
Such an approach has served as a basis for further studies of 1+2 QuIC \cite{Atanasov96, Rioux11}, and in this article we follow it to provide a first step in elucidating 2+3 QuIC as well. 
We derive expressions for the optical injection coefficients at the initial time when the fields are incident, and evaluate them for different stoichiometries of AlGaAs using a 30-band $\bm{k}\cdot\bm{p}$ model.
We compute all the symmetry-allowed injection coefficients corresponding to different polarizations of the incident fields, and analyze their frequency dependence over a range where the injection of carriers that do not contribute to the current is suppressed.  
That is, considering 3-photon absorption of photons at energy $\hbar\omega$ and 2-photon absorption of photons at energy $3\hbar\omega/2$, we require $2\hbar\omega$ to be less than the band gap. 
The alloy AlGaAs is an ideal material for 2+3 QuIC, as its stoichiometry can be chosen to yield a band gap appropriate for the available laser wavelengths. 
Experiments demonstrating 2+3 QuIC of photocurrents in AlGaAs are being reported in another article \cite{Wang17}.

As would be expected, there are qualitative differences between 2+3 QuIC and 1+2 QuIC. 
For instance, there is a change of sign in the current injection coefficient for different frequencies, which is due to an interplay between intraband and interband processes contributing to 3-photon absorption.
We also find that 2+3 QuIC leads to sharper distributions of carriers in the BZ than 1+2 QuIC. 
The sharper distribution of carriers leads to a higher swarm velocity, which is a desirable feature for photocurrents, and it also opens the possibility of exciting carriers in semiconductors in a tailored fashion.

The outline of this article is the following: In Sec.~\ref{sec:response}
we present a method to compute the optical injection
rates for a generic material. 
In Sec.~\ref{sec:AlGaAs} we describe
the model used for AlGaAs. In Sec.~\ref{sec:results} we present
our results for carrier and current injection from 2- and 3-photon
absorption (2PA and 3PA) processes in AlGaAs. 
 We also discuss the efficiency of the current
injection by analyzing
the swarm velocity, and computing the optimal laser intensities. In Sec.~\ref{sec:conclusion} we discuss
the implications of our results and present our conclusions.
We list the independent components of the optical injection tensors for zincblende lattice
symmetry in the Appendix.

\section{Optical injection rates }

\label{sec:response}

In this article, we restrict ourselves to computing the optical injection rates at the initial time when the fields are incident. We neglect scattering processes, carrier acceleration in the BZ, and several other effects that later influence the dynamics of the injected carriers.   
We thus use a Fermi Golden Rule approach \citep{Rioux12,Muniz14} as it is adequate for computing transition rates. Other approaches based on solving equations of motion \cite{Nastos07,Sternemann13,Sternemann16} for the system have been used for 1+2 QuIC, and they can be straightforwardly extended to 2+3 QuIC.

Assuming the independent particle approximation, we consider a system
described by a Hamiltonian $\mathcal{H}_{0}$ in the absence of any
external perturbation, so the full Hamiltonian $\mathcal{H}\left(t\right)$
in the presence of the external perturbation $\mathcal{V}_{ext}\left(t\right)$
is $\mathcal{H}\left(t\right)=\mathcal{H}_{0}+\mathcal{V}_{ext}\left(t\right)$,
where in the basis of eigenstates of $\mathcal{H}_{0}$,  
\begin{align}
\mathcal{H}_{0}= & \sum_{n\bm{k}}\hbar\omega_{n\bm{k}}a_{n\bm{k}}^{\dagger}a_{n\bm{k}},\\
\mathcal{V}_{{\rm ext}}\left(t\right)= & \sum_{mn\bm{k}}a_{m\bm{k}}^{\dagger} V_{mn\bm{k}}\left(t\right)a_{n\bm{k}},
\end{align}
where $\left|n \bm{k}\right\rangle = a_{n\bm{k}}^{\dagger} |{\rm vac}\rangle $ indicates a Bloch state corresponding to band $n$, with
crystal momentum $\bm{k}$, and energy $\hbar \omega_{n \bm{k}}$.
In the interaction picture, the creation and annihilation fermion
operators are $a_{n\bm{k}}^{\dagger}\left(t\right)=a_{n\bm{k}}^{\dagger}e^{i\omega_{n}t}$
and $a_{n\bm{k}}\left(t\right)=a_{n\bm{k}}e^{-i\omega_{n}t}$, the external perturbation operator is 
$\mathcal{V}_{\rm I} \left( t \right) = e^{i \mathcal{H}_{0} t/\hbar} \mathcal{V}_{\rm ext}\left(t\right) e^{-i \mathcal{H}_{0} t/\hbar}$,
and the time-evolution operator can be expanded as 
\begin{equation}
\mathcal{U}\left(t\right)=1+\sum_{N=1}^{\infty}\int_{-\infty}^{t}\frac{dt_{N}}{i\hbar}\mathcal{V}_{\rm I}\left(t_{N}\right)\dots\int_{-\infty}^{t_{2}}\frac{dt_{1}}{i\hbar}\mathcal{V}_{\rm I}\left(t_{1}\right),
\end{equation}
and the terms of each order in $\mathcal{V}_{{\rm ext}}$ can be obtained
from the previous one by 
\begin{align}
\mathcal{U}_{N}\left(t\right)= & \int_{-\infty}^{t}\frac{dt_{N}}{i\hbar}\sum_{mn\bm{k}}\nonumber \\
 & \>\quad a_{m\bm{k}}^{\dagger}\left(t_{N}\right) V_{mn\bm{k}}\left(t_{N}\right) a_{n\bm{k}}\left(t_{N}\right) \mathcal{U}_{N-1}\left(t_{N}\right),\label{eq:Uexpan}
\end{align}
where $\omega_{mn\bm{k}}=\omega_{m\bm{k}}-\omega_{n\bm{k}}$,
and $\mathcal{U}_{0}\left(t\right)=1$. 
We are interested in the excitation
of an electron from a valence band $v$ to a conduction band $c$
due to the external field. 
This excited state is $\left|cv\bm{k}\right\rangle = a_{c\bm{k}}^{\dagger}  a_{v\bm{k}} |{\rm gs}\rangle $,
where $|{\rm gs}\rangle $ is the eigenstate of $\mathcal{H}_{0}$
with filled valence bands. The state of the system is described by 
\begin{align}
\left|\psi\right\rangle = & \mathcal{U}\left(t\right) | {\rm gs} \rangle \nonumber \\
= & \gamma_{0}|{\rm gs}\rangle +\sum_{cv\bm{k}}\gamma_{cv\bm{k}}\left(t\right) \left|cv\bm{k}\right\rangle +\dots
\label{eq:psi}
\end{align}
where the coefficient 
\begin{equation}
\gamma_{cv\bm{k}}\left(t\right) = \left\langle cv\bm{k} \right| \mathcal{U} \left(t\right) \left|{\rm gs}\right\rangle 
\end{equation}
indicates the degree to which the system has been excited into the $\left|cv\bm{k}\right\rangle$ state, so they allow us to compute injection rates. 
We point out that the $\gamma_{cv\bm{k}}\left(t\right)$ coefficients are related to a single-particle density matrix $\rho_{mn}\left(t\right) = \left\langle \psi \right| a_{m\bm{k}}^{\dagger}\left(t\right)  a_{n\bm{k}}\left(t\right) \left| \psi \right\rangle$, which could be used for computing $\chi^{(5)}$, but that is a much more complicated calculation and it includes several non-resonant effects that are not the focus of our study. Thus we use a simpler parametrization for the states of the system in terms of  $\gamma_{cv\bm{k}}\left(t\right)$.  

For a full Hamiltonian $\mathcal{H}\left(t\right)$ that follows from a Hamiltonian for a single particle of the form
\begin{align}
\mathscr{H} \left(\bm{x},\bm{p};t\right)= & \frac{1}{2m}\left(\bm{p}-e\bm{A}\left(t\right)\right)^{2} \nonumber \\ 
& \> + \mathscr{H}_{SO}\left(\bm{x},\bm{p}-e\bm{A}\left(t\right)\right) + \mathscr{V}_{{\rm lat}}\left(\bm{x}\right),\label{eq:single_particle}
\end{align} 
 where $\bm{x}$ and $\bm{p}$ are position and momentum
operators, $H_{SO}$ is the spin-orbit term, and $\mathscr{V}_{{\rm lat}}\left(\bm{x}\right)$
is the lattice potential energy. 
Here we neglect a contribution to the interaction
that is solely a function of time ($\sim\left[ A\left(t\right) \right]^{2}$),
for it will not lead to any transitions, and we work in a gauge where
the electric field $\bm{E}\left(t\right)$, assumed independent
of position, is fully described by the vector potential $\bm{A}\left(t\right)$. 
We point out that for a general $\bm{A}$ dependent on the position, neglecting the $\left[ A \right]^{2}$ term in the Hamiltonian would be problematic. 
The interaction term in the Hamiltonian takes the form
$\mathscr{V}_{{\rm ext}}\left(t\right)  =-e\bm{\mathfrak{v}}\cdot\bm{A}\left(t\right),$
where $e=-\left|e\right|$ is the charge of the electron and $\bm{\mathfrak{v}} = -e^{-1} \partial \mathscr{H} / \partial \bm{A} $
is the velocity operator. 
Indeed, the interaction is of the form we consider for any unperturbed
Hamiltonian for a single particle that is at most quadratic in the momentum. 

We take the vector potential to be 
\begin{equation}
\bm{A}\left(t\right)=\sum_{\alpha}\bm{A}_{\alpha}e^{-i\left(\omega_{\alpha}+i\epsilon\right)t}=-\sum_{\alpha}\frac{i}{\omega_{\alpha}}\bm{E}_{\alpha}e^{-i\left(\omega_{\alpha}+i\epsilon\right)t},
\end{equation}
with $\omega_{\alpha}=\pm\omega,\,\pm3\omega/2$; here $\epsilon\to0^{+}$
describes turning on the field from $t=-\infty$. The $\gamma_{cv\bm{k}}\left(t\right)$
coefficients can be expanded as $\gamma_{cv\bm{k}}^{\left(N\right)}\left(t\right)=\left\langle cv\bm{k}\right|\mathcal{U}_{N}\left(t\right)|{\rm gs}\rangle $
following the expansion \eqref{eq:Uexpan} of $\mathcal{U}\left(t\right)$
for an incident optical field, so we can write the coefficients $\gamma_{cv\bm{k}}^{\left(N\right)}\left(t\right)$
as 
\begin{equation}
\gamma_{cv\bm{k}}^{\left(N\right)}\left(t\right)=\mathscr{R}_{cv\bm{k}}^{\left(N\right)}\frac{e^{-i\left(\Omega_{N}-\omega_{cv\bm{k}}+i\epsilon\right)t}}{\Omega_{N}-\omega_{cv\bm{k}}+i\epsilon},\label{eq:amp_gral}
\end{equation}
where $\Omega_{N}=\omega_{1}+\ldots+\omega_{N}$. The coefficients
$\mathscr{R}_{cv\bm{k}}^{\left(N\right)}$ involve the electric
field amplitudes $\bm{E}_{\alpha}$ according to \begin{equation}
\mathscr{R}_{cv\bm{k}}^{\left(N\right)} = R_{cv\bm{k}}^{\left(N\right)a\ldots b}\left(\omega_{\alpha},\dots,\omega_{\beta}\right)E_{\alpha}^{a}\ldots E_{\beta}^{b},
\label{eq:Rcoefs}
\end{equation}
where repeated indices are summed; here superscripts refer to Cartesian
indices and subscripts to incident frequency components. For the lower
orders we have 
\begin{align}
R_{cv\bm{k}}^{\left(1\right)a}\left(\omega_{\alpha}\right)= & \sum_{\alpha}\dfrac{ie}{\hbar\omega_{\alpha}}v_{cv\bm{k}}^{a}, \nonumber \\
R_{cv\bm{k}}^{\left(2\right)ab}\left(\omega_{\alpha},\omega_{\beta}\right)= & \sum_{\alpha\beta}\dfrac{-e^{2}}{\hbar^{2}\omega_{\alpha}\omega_{\beta}} \nonumber \\   
& \> \times  \left(\sum_{c^{\prime}}\dfrac{v_{cc^{\prime}\bm{k}}^{a}v_{c^{\prime}v\bm{k}}^{b}}{\omega_{\beta}-\omega_{c^{\prime}v\bm{k}}}-\sum_{v^{\prime}}\dfrac{v_{cv^{\prime}\bm{k}}^{b}v_{v^{\prime}v\bm{k}}^{a}}{\omega_{\beta}-\omega_{cv^{\prime}\bm{k}}}\right),\label{eq:R2coef}
\end{align}
and 
\begin{widetext}
\begin{align}
R_{cv\bm{k}}^{\left(3\right)abd}\left(\omega_{\alpha},\omega_{\beta},\omega_{\delta}\right)=\sum_{\alpha\beta\gamma}\dfrac{ie^{3}}{\hbar^{3}\omega_{\alpha}\omega_{\beta}\omega_{\delta}} & \left[\sum_{c^{\prime}}\dfrac{v_{cc^{\prime}\bm{k}}^{a}}{\omega_{\alpha}-\omega_{cc^{\prime}\bm{k}}}\left(\sum_{c^{\prime\prime}}\dfrac{v_{c^{\prime}c^{\prime\prime}\bm{k}}^{b}v_{c^{\prime\prime}v\bm{k}}^{d}}{\omega_{\delta}-\omega_{c^{\prime\prime}v\bm{k}}}-\sum_{v^{\prime}}\dfrac{v_{c^{\prime}v^{\prime}\bm{k}}^{d}v_{v^{\prime}v\bm{k}}^{b}}{\omega_{\delta}-\omega_{c^{\prime}v^{\prime}\bm{k}}}\right)\right.\nonumber \\
 & \>\quad-\sum_{v^{\prime}}\left(\sum_{c^{\prime}}\dfrac{v_{cc^{\prime}\bm{k}}^{b}v_{c^{\prime}v^{\prime}\bm{k}}^{d}}{\omega_{\delta}-\omega_{c^{\prime}v^{\prime}\bm{k}}}-\sum_{v^{\prime\prime}}\dfrac{v_{cv^{\prime\prime}\bm{k}}^{d}v_{v^{\prime\prime}v^{\prime}\bm{k}}^{b}}{\omega_{\delta}-\omega_{cv^{\prime\prime}\bm{k}}}\right)\dfrac{v_{v^{\prime}v\bm{k}}^{a}}{\omega_{\alpha}-\omega_{v^{\prime}v\bm{k}}}\nonumber \\
 & \>\left.\quad-\sum_{c^{\prime}v^{\prime}}\left(\dfrac{v_{cv^{\prime}\bm{k}}^{b}v_{v^{\prime}c^{\prime}\bm{k}}^{a}v_{c^{\prime}v\bm{k}}^{d}}{\left(\omega_{\alpha}-\omega_{v^{\prime}c^{\prime}\bm{k}}\right)\left(\omega_{\delta}-\omega_{c^{\prime}v\bm{k}}\right)}+\dfrac{v_{cv^{\prime}\bm{k}}^{d}v_{v^{\prime}c^{\prime}\bm{k}}^{a}v_{c^{\prime}v\bm{k}}^{b}}{\left(\omega_{\delta}-\omega_{cv^{\prime}\bm{k}}\right)\left(\omega_{\alpha}-\omega_{v^{\prime}c^{\prime}\bm{k}}\right)}\right)\right].\label{eq:R3coef}
\end{align}
\end{widetext}

The expectation value of the density $\left\langle M\right\rangle $ of a generic quantity associated with an operator
$\mathcal{\mathcal{M}}=\underset{mn\bm{k}}{\sum}a_{m\bm{k}}^{\dagger}\left(t\right) M_{mn\bm{k}} a_{n\bm{k}}\left(t\right)$
can be computed from Eq.~\eqref{eq:psi}, and it will have terms independent, linear, and quadratic on $\gamma_{cv\bm{k}} \left(t\right)$. The independent term corresponds to the expectation value $\left\langle M \right\rangle $ in the absence of perturbations. 
 The linear terms have accompanying $e^{\pm i\omega_{cv\bm{k}}t}$ factors, thus they are fast oscillating and we ignore them as we are interested in computing injection rates. The quadratic term in $\gamma_{cv\bm{k}} \left(t\right) $ then gives us the injection rates, and it is 
\begin{align}
\Delta \left\langle M\right\rangle =\frac{1}{L^{D}} & \sum_{cvc^{\prime}v^{\prime}\bm{k}}  \left\langle c^{\prime}v^{\prime}\bm{k} \right|M\left|cv\bm{k}\right\rangle \nonumber \\
 & \>\quad\times \gamma_{c^{\prime}v^{\prime}\bm{k}}^{\ast}\left(t\right) \gamma_{cv\bm{k}}\left(t\right) e^{i\omega_{c^{\prime}v^{\prime}\bm{k}}t}e^{-i\omega_{cv\bm{k}}t}  \nonumber \\
=\frac{1}{L^{D}} & \sum_{cvc^{\prime}v^{\prime}\bm{k}}\left(M_{c^{\prime}c\bm{k}}\delta_{v^{\prime}v}-M_{v^{\prime}v\bm{k}}\delta_{c^{\prime}c}\right)\nonumber \\
 & \>\quad\times \gamma_{c^{\prime}v^{\prime}\bm{k}}^{\ast}\left(t\right) \gamma_{cv\bm{k}}\left(t\right) e^{i\omega_{c^{\prime}v^{\prime}\bm{k}}t}e^{-i\omega_{cv\bm{k}}t}, 
\label{eq:avgM}
\end{align}
where $L$ is a normalization length, $D$ is the spatial dimension of the system, 
and $\Delta \left\langle M\right\rangle = \left\langle M\right\rangle - \left\langle M\right\rangle _0$ indicates the change to the expectation value $\left\langle M\right\rangle$ due to the perturbation \cite{Mahon18}.
Since we are interested in the non-oscillatory response of the system, so
we focus on the $\Omega_{N}=\Omega_{N^{\prime}} =\Omega$ contributions to
Eq.~\eqref{eq:avgM}. 
To compute the injection rate $d\left\langle M\right\rangle /dt$
associated with Eq.~\eqref{eq:avgM}, it is important to realize that 
\begin{align}
 & \frac{d}{dt}\left. \left(\gamma_{c^{\prime}v^{\prime}\bm{k}}^{\ast}\left(t\right)  \gamma_{cv\bm{k}}\left(t\right) e^{i\omega_{c^{\prime}v^{\prime}\bm{k}}t}e^{-i\omega_{cv\bm{k}}t}\right)\right|_{t \to 0, \epsilon \to 0}
\nonumber \\ 
 & = \sum_{N, N^{\prime} } 
 \left. \frac{ \mathscr{R}_{c^{\prime}v^{\prime}\boldsymbol{k}}^{\left( N^{\prime} \right)\ast} \mathscr{R}_{cv\boldsymbol{k}}^{\left( N \right)} 
 \times 2\epsilon}{\left(\Omega-\omega_{c^{\prime}v^{\prime}\bm{k} }-i\epsilon\right)\left(\Omega-\omega_{cv\bm{k} }+i\epsilon\right)}\right|_{\epsilon \to 0} 
\nonumber \\
 & =\sum_{N, N^{\prime} } \left. 
 \mathscr{R}_{c^{\prime}v^{\prime}\bm{k}}^{\left(N^{\prime}\right)\ast} \mathscr{R}_{cv\bm{k}}^{\left(N\right)} 2\pi\delta\left(\Omega -\omega_{cv\bm{k} }\right)
 \right|_{\omega_{cv\bm{k} }=\omega_{c^{\prime}v^{\prime}\bm{k} }} ,
\label{eq:Fermi}
\end{align}
where the sums over $N$ and $N^{\prime}$ are restricted to the cases when $\Omega_{N}=\Omega_{N^{\prime}} =\Omega$. 
The fact that the $\mathscr{R}_{cv\bm{k}}^{\left(N\right)}$
coefficients are always accompanied by $\delta\left(\Omega -\omega_{cv\bm{k} }\right)$
in the expression for the response allows for substitutions $3\hbar\omega-\omega_{cv\bm{k} }=0$
that were used to write $R_{cv\bm{k}}^{\left(3\right)abd}$
in a simpler way in Eq.~\eqref{eq:R3coef}. 
The resonance described by Eq.~\eqref{eq:Fermi} corresponds to a divergence in the electric susceptibility $\chi^{(5)}$, which also describes many non-resonant effects that we are ignoring in this article, as they are much weaker in comparison to resonant ones.

The injection rate of the density of a generic quantity $\left\langle M\right\rangle $ can be obtained by taking a time derivative of Eq.~\eqref{eq:avgM}. Using Eq.~\eqref{eq:Fermi} and the dependence of 
$\mathscr{R}_{c^{\prime}v^{\prime}\bm{k}}^{\left(N^{\prime}\right)\ast}\mathscr{R}_{cv\bm{k}}^{\left(N\right)}$ 
given by Eq.~\eqref{eq:Rcoefs}, we can write the 
injection rate of $\left\langle M\right\rangle $ due to the interference of an $N^{\prime}$ photon process with an $N$ photon process in terms of a coefficient, 
\begin{equation}
\frac{d}{dt}\left\langle M\right\rangle =\mu^{abd\ldots,pq\ldots}\left(\Omega\right)E_{-\alpha}^{a}E_{-\beta}^{b}E_{-\delta}^{d}\ldots E_{\rho}^{p}E_{\sigma}^{q}\ldots+c.c.,\label{eq:injectM}
\end{equation}
where there are $N^{\prime}$ frequency labels ($\alpha,\beta,\delta,...)$
and $N$ frequency labels ($\rho,\sigma,...$), and $\Omega_{N}=\Omega_{N^{\prime}} =\Omega$. 
The injection rate coefficient $\mu^{abd\dots,pq\dots}\left(\Omega\right)$ is assembled
from the matrix elements $M_{ab \bm{k}}$ in  Eq.~\eqref{eq:avgM} and coefficients $R_{cv\bm{k}}^{\left(N\right) pq\dots}$ in Eq.~\eqref{eq:Fermi}. 
Taking the continuous momentum limit, we have  
\begin{align}
\mu^{abd\dots,pq\dots}\left(\Omega\right) = & 2\pi\int\frac{d\bm{k}}{\left(2\pi\right)^{D}} 
\sum_{cvc^{\prime}v^{\prime}} 
\nonumber \\ 
&  \left(M_{c^{\prime}c\bm{k}}\delta_{v^{\prime}v}-M_{v^{\prime}v\bm{k}}\delta_{c^{\prime}c}\right) 
\delta_{\omega_{cv\bm{k} }=\omega_{c^{\prime}v^{\prime}\bm{k} }}
 \nonumber \\
 & \>\times R_{c^{\prime}v^{\prime}\bm{k}}^{\left( N^{\prime} \right)abd \dots \ast}R_{cv\bm{k}}^{\left( N \right)pq \dots}\delta\left(\Omega-\omega_{cv\bm{k} }\right).
\label{eq:coeffMN}
\end{align}
We will use instances of $\mu^{abd\dots,pq\dots}\left(\Omega\right)$ for carrier and current density in this article.

\subsection*{Quantum interference of 2- and 3-photon processes }

\begin{figure}[hbt!]
\includegraphics[width=0.6\columnwidth]{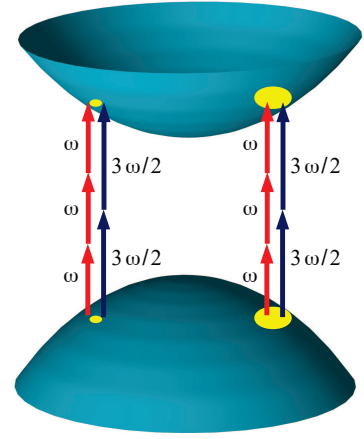}
\caption[]{Depiction of 2+3 QuIC showing the destructive (left) and constructive
(right) interference in different regions of the Brillouin zone. }
\label{fig:QuIC23} 
\end{figure}

The processes of 3PA with frequency $\omega$ and 2PA with frequency
$3\omega/2$ can interfere since the total frequency for each of them
is $\Omega=3\omega$, as illustrated in Fig.~\ref{fig:QuIC23}. For such processes the frequencies are all equal
in the equations \eqref{eq:R2coef} and \eqref{eq:R3coef} for the
coefficients $R_{cv\bm{k}}^{\left(2\right)}$ and $R_{cv\bm{k}}^{\left(3\right)}$,
and symmetrizing their components leads to some simplifications. Using
$\omega_{\beta}=\omega_{cv\bm{k}}-\omega_{\alpha}$ and $\omega_{\alpha}=\omega_{\beta}=3\omega/2$
in Eq.~\eqref{eq:R2coef}, the second order coefficient simplifies to   
\begin{equation}
R_{cv\bm{k}}^{\left(2\right)ab}\left(\frac{3\omega}{2},\frac{3\omega}{2}\right)=\frac{-4e^{2}}{9\hbar^{2}\omega^{2}}\sum_{m}\dfrac{v_{cm\bm{k}}^{a}v_{mv\bm{k}}^{b}}{\left(\frac{3\omega}{2}-\omega_{mv\bm{k}}\right)},\label{eq:R2coefSimp}
\end{equation}
and using $\omega_{\beta}+\omega_{\gamma}=\omega_{cv\bm{k}}-\omega_{\alpha}$
and $\omega_{\alpha}=\omega_{\beta}=\omega_{\gamma}=\omega$ in Eq.~\eqref{eq:R3coef}, the third order coefficient simplifies to 
\begin{equation}
R_{cv\bm{k}}^{\left(3\right)abd}\left(\omega,\omega,\omega\right)=\frac{ie^{3}}{\hbar^{3}\omega^{3}}\sum_{mn}\dfrac{v_{cm\bm{k}}^{a}v_{mn\bm{k}}^{b}v_{nv\bm{k}}^{d}}{\left(\omega-\omega_{cm\bm{k}}\right)\left(\omega-\omega_{nv\bm{k}}\right)}.\label{eq:R3coefSimp}
\end{equation}
Notice that the denominators in Eqs.~\eqref{eq:R2coefSimp} and \eqref{eq:R3coefSimp} 
are minimal for $m,n=c,v$, so the dominant contributions  to $R_{cv\bm{k}}^{\left(2\right)}$
always involve intraband velocity matrix elements, while  $R_{cv\bm{k}}^{\left(3\right)}$
also has contributions from interband velocity matrix elements 
\footnote{The denominators in Eqs.~\eqref{eq:R2coefSimp} and \eqref{eq:R3coefSimp}
do not lead to any divergences because of the assumption that $2\hbar\omega$
is below the gap.}. 
Intraband velocity matrix elements are associated with the corresponding band dispersion,
$v_{nn\bm{k}}^{a}=\partial_{\bm{k}}^{a}\omega_{n\bm{k}}$,
which vanishes at the $\bm{k}$ point corresponding to the bandgap.
Thus $R_{cv\bm{k}}^{\left(2\right)}$ is zero for total photon
energies corresponding to the band gap, and increases for larger excess
photon energies. The dependence of $R_{cv\bm{k}}^{\left(3\right)}$
on the total photon energy is different, as it depends on both interband
and intraband velocity matrix elements. For total photon energies
just above the gap, $R_{cv\bm{k}}^{\left(3\right)}$ is determined
mainly by the interband matrix elements, but as the photon excess
energy increases $R_{cv\bm{k}}^{\left(3\right)}$ becomes
dominated by the intraband matrix elements, since the electronic transitions
occur at $\bm{k}$ points with larger band dispersion.

The injection rate coefficients corresponding to the interference
of 2- and 3-photon processes can then be computed from Eqs.~\eqref{eq:coeffMN} as 
\begin{align}
\mu_{2+3}^{abd,fg}\left(\Omega\right)= & 2\pi\int\frac{d\bm{k}}{\left(2\pi\right)^{D}}\sum_{cvc^{\prime}v^{\prime}}\left(M_{c^{\prime}c\bm{k}}\delta_{v^{\prime}v}-M_{v^{\prime}v\bm{k}}\delta_{c^{\prime}c}\right)\nonumber \\
 & \>\times\delta_{\omega_{cv\bm{k}}=\omega_{c^{\prime}v^{\prime}\bm{k} }}R_{c^{\prime}v^{\prime}\bm{k}}^{\left(3\right)abd\ast}R_{cv\bm{k}}^{\left(2\right)fg}\delta\left(\Omega-\omega_{cv\bm{k} }\right).
\label{eq:coeff}
\end{align}
For the plots in the next sections we use a frequency broadening $\Delta$ corresponding to
$\hbar\Delta=13{\rm m}e$V.

The factor $R_{c^{\prime}v^{\prime}\bm{k}}^{\left(3\right)abd\ast}R_{cv\bm{k}}^{\left(2\right)fg}$
changes sign under a transformation $\bm{k}\to-\bm{k}$, resulting in constructive versus destructive interference in opposite points of the Brillouin zone. 
In Fig.~\ref{fig:QuIC23} we illustrate constructive versus destructive interference of 2- and
3-photon processes at opposite points in the Brillouin zone. 

\section{Electronic model of ${\rm {\bf AlGaAs}}$ }

\label{sec:AlGaAs}

We use a 30-band $\bm{k}\cdot\bm{p}$ model for computing
the electronic bands. The model has free parameters associated with
energies and momentum matrix elements at the $\Gamma$ point, and
the parameters are adjusted to match the experimental results for
band energies from $-5\,e$V to $4\,e$V, such that computations of
optical absorption coefficients are expected to be reliable for photon
energies up to $6\,e$V.

\begin{figure}[hbt!]
\includegraphics[width=0.9\columnwidth]{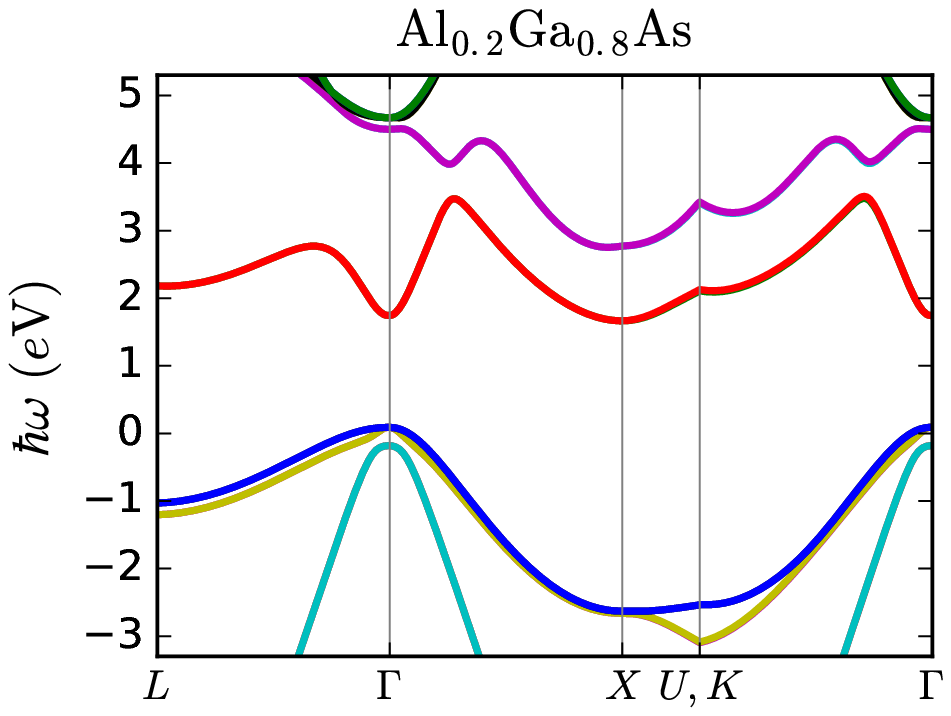}
\includegraphics[width=0.9\columnwidth]{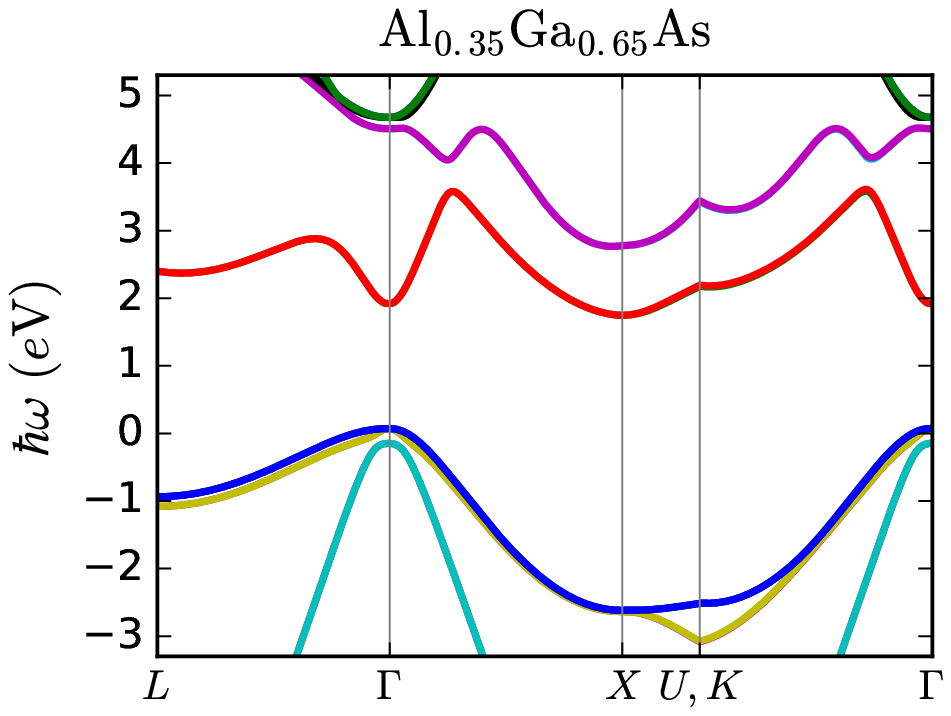}
\caption[]{ Electronic bandstructure of AlGaAs for two stoichiometries. 
For all the stoichiometries $\alpha$ in the range that we consider, $0.18\lesssim\alpha\lesssim0.38$, the band structures are vere similar, and their main difference is their  bandgap.}
\label{fig:bands} 
\end{figure}

Using the $\Gamma$ point as the expansion point for a $\bm{k}\cdot\bm{p}$
model, the effective Hamiltonian that acts only on the periodic part
of an energy eigenfunction of crystal momentum $\bm{k}$ is
\begin{equation}
H_{{\rm eff}}=H+\frac{\hbar}{m}\bm{k}\cdot\bm{p}+\frac{\hbar^{2}k^{2}}{2m},
\end{equation}
where $H$ is the Hamiltonian (\ref{eq:single_particle}) with the
vector potential set equal to zero; in this model \cite{richard04}
we neglect the $\bm{k}$ dependence of the effective spin-orbit
term. The second term on the right-hand-side is the usual $\bm{k}\cdot\bm{p}$
contribution, and the last term is the contribution to the kinetic
energy only due to the lattice momentum. The basis of states has 8
sets \cite{richard04}, 4 of them corresponding to the $\Gamma_{1}$
representation of the point group $T_{d}$ (or $43m$), 3 corresponding
to the $\Gamma_{4}$ representation, and 1 to the $\Gamma_{3}$ representation.
The $\Gamma_{1}$ representation has only 1 state, $\Gamma_{4}$ has
3 states, and $\Gamma_{3}$ has 2 states, so in total we have $4\times1+3\times3+1\times2=15$
states before considering spin; we denote these states as $\left|A\right\rangle $,
$\left|B\right\rangle $, etc. Tensor products of these are taken
with spin states to get 30 states in all. Terms $\left\langle A\right|H_{{\rm eff}} \left|B\right\rangle$
are then $2\times2$ matrices, and take the form 
\begin{align}
\left\langle A\right|H_{{\rm eff}}\left|B\right\rangle = & E_{A}\delta_{AB}\sigma_{0} +\frac{i}{3}\bm{\Delta}_{AB}\cdot\bm{\sigma} \nonumber \\
&  +i\bm{P}_{AB}\cdot\bm{k}\sigma_{0}+\frac{\hbar^{2}k^{2}}{2m}\delta_{AB}\sigma_{0},
\end{align}
where $\sigma_{0}$ is the unit $2\times2$ matrix and the components
of $\bm{\sigma}$ are the usual Pauli matrices. The free parameters
of the model are the energies $E_{A}$, the matrix elements of the
spin-orbit term $\Delta_{AB}$, and the matrix elements of the momentum
operator $\bm{P}_{AB}$. Since the basis for the states is
the same at every $\bm{k}$ point \cite{Salazar17}, the corresponding
$2\times2$ matrices corresponding to the velocity operator $\left\langle A|\bm{\mathfrak{v}}|B\right\rangle $
are diagonal in the spin sector, 
\begin{align}
\left\langle A\right|\mathfrak{v}^{a}\left|B\right\rangle 
= & \frac{1}{\hbar}\frac{\partial}{\partial k^{a}}\left\langle A\right|H_{\rm eff}\left|B\right\rangle  \nonumber \\ 
= & \left(\frac{i}{\hbar}P_{AB}^{a}+\frac{\hbar k^{a}}{m}\delta_{AB}\right)\sigma_{0},
\end{align}
 from which the matrix elements of the velocity operator between the
energy eigenstates can be determined. 

\begin{figure}[hbt!]
\includegraphics[width=0.9\columnwidth]{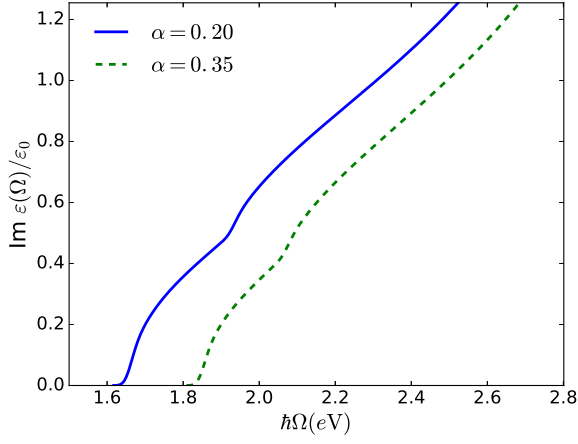}
\caption[]{Imaginary part of the dielectric function for two stoichiometries $\alpha$. }
\label{fig:dielec}
\end{figure}

For GaAs \cite{richard04} and AlAs \cite{fraj07} we use reported
parameters adjusted for room temperature, while the parameters for
Al$_{\alpha}$Ga$_{1-\alpha}$As are obtained from a linear interpolation
according to the stoichiometry. This approximation is accurate within
an energy tolerance corresponding to room temperature 
\footnote{The corrections that are quadratic on the stoichiometry parameter
$\alpha$ are small, and do not lead to significant changes in the
band energies within a tolerance given by room temperature. }. 
The chosen parameters lead to effective masses, $g$-factors, and Luttinger parameters that
are in good agreement with experimental data \cite{richard04,Rioux11th,fraj07}. More important for the
problems we consider, the band structures and linear optical absorption
spectra are also in good agreement with experimental data. In Fig.~\ref{fig:bands}
we show the relevant electronic bands for two different stoichiometries,
and in Fig.~\ref{fig:dielec} we show the imaginary parts of the
corresponding dielectric functions, which are related to the 1-photon
absorption rates (or carrier injection) by ${\rm Im} ~\varepsilon\left(\Omega\right)=\hbar\xi^{xx}\left(\Omega\right)/2\epsilon_{0}$.
Calculations for 1+2 QuIC in GaAs and Si have  
been computed \cite{Nastos07} using the same $\bm{k}\cdot\bm{p}$ model and Density Functional Theory for comparison. 
For GaAs there is good agreement between the two models even for the non-diagonal spin injection tensors, and for Si the discrepancy between the two models is significant only at high frequencies. 
For GaAs the $\bm{k}\cdot\bm{p}$ model for the full BZ provides a good agreement with the Local Density Approximation, and for the frequencies we consider in 2+3 QuIC we are restricted to the region of the BZ around the $\Gamma$ point, where even less sophisticated $\bm{k}\cdot\bm{p}$ models give good results.

As already mentioned, the 30-band model allows us to perform  reliable calculations for total photon energies up to $6\,e$V. 
However, for the energies we are most interested -- below $3\,e$V -- we can get accurate results for the optical absorption coefficients even if only 6 valence and 2 conduction bands are included in the model, and only the valley including the $\Gamma$ point of the BZ are considered. 
We also point out that the model we use is applicable to most zincblende semiconductors, where each specific material corresponds to a particular set of parameters. 
Since the bandstructures of zincblende semiconductors are qualitatively similar, our results presented in the next sections are qualitatively valid for most direct-gap zincblende semiconductors.  

\section{Quantum interference control using two- and three-photon absorption in ${\rm {\bf AlGaAs}}$ }

\label{sec:results}

We consider two incident fields of different frequencies with amplitudes
$\bm{E}_{\omega}=E_{\omega}e^{i\phi_{\omega}}\hat{\bm{e}}_{\omega}$
and $\bm{E}_{3\omega/2}=E_{3\omega/2}e^{i\phi_{3\omega/2}}\hat{\bm{e}}_{3\omega/2}$,
where $E_{\omega}>0$ and $E_{3\omega/2}>0$ are the field magnitudes,
the unit vectors $\hat{\bm{e}}_{\omega}$ and $\hat{\bm{e}}_{3\omega/2}$
indicate their polarizations, and $\phi_{\omega}$ and $\phi_{3\omega/2}$
indicate their phases. We also define the phase parameter $\Delta\phi=2\phi_{3\omega/2}-3\phi_{\omega}$,
which will be useful later. 
We assume that the field at $3\omega/2$ 
has a weaker intensity than the field at $\omega$,
and we demand that the frequencies satisfy $2\omega<\Delta_{g}<3\omega$,
where $E_{g}=\hbar\Delta_{g}$ is the optical gap. Therefore only
3PA processes are important for the lower frequency field $\bm{E}_{\omega}$,
while only 2PA processes are relevant for the higher frequency field
$\bm{E}_{3\omega/2}$; the 3PA associated with $\bm{E}_{3\omega/2}$
is weaker due to the lower intensity of the field, and we neglect
it.

We focus on Al concentrations $\alpha$ such that $0.18\lesssim\alpha\lesssim0.38$,
since Al$_{\alpha}$Ga$_{1-\alpha}$As with $\alpha$ too small has
a band gap smaller than $2\hbar\omega$ for telecommunication wavelengths
($\hbar\omega\sim0.8\,e$V), and Al$_{\alpha}$Ga$_{1-\alpha}$As
with $\alpha$ too large is too reactive. 

\begin{figure}[hbt!]
\includegraphics[width=0.9\columnwidth]{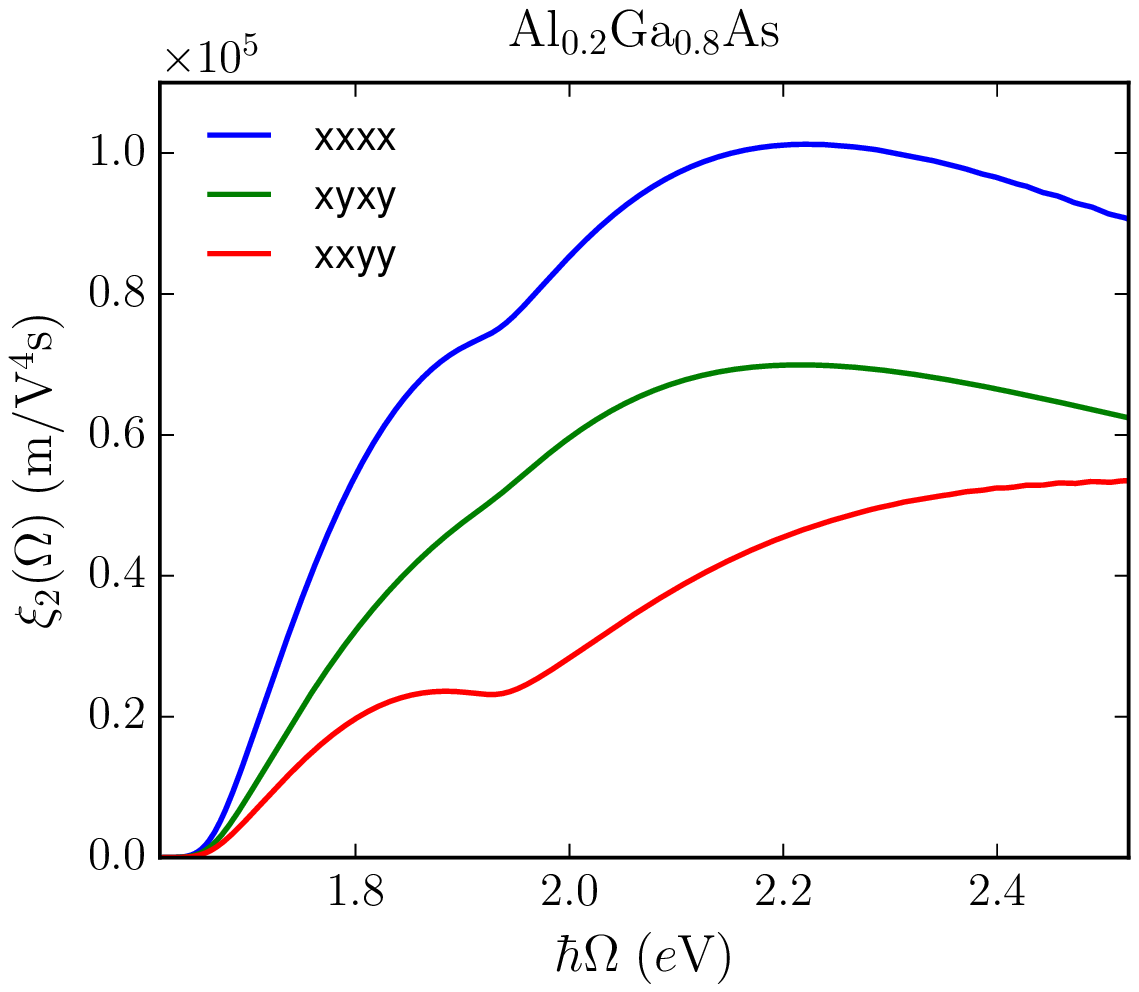}
\includegraphics[width=0.9\columnwidth]{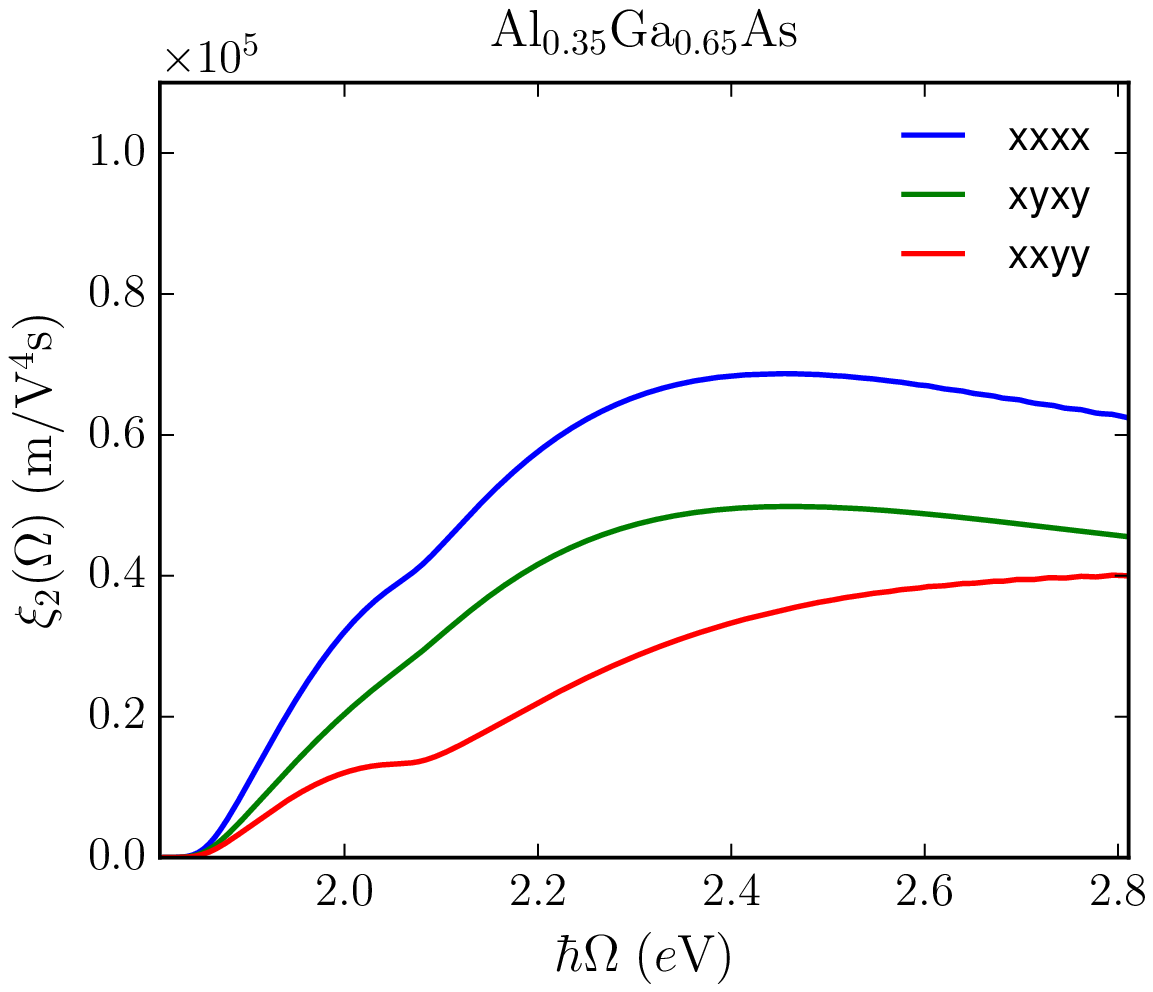}
\caption[]{ Two-photon carrier injection coefficients for two stoichiometries. }
\label{fig:carrier2} 
\end{figure}

\begin{figure}[hbt!]
\includegraphics[width=0.9\columnwidth]{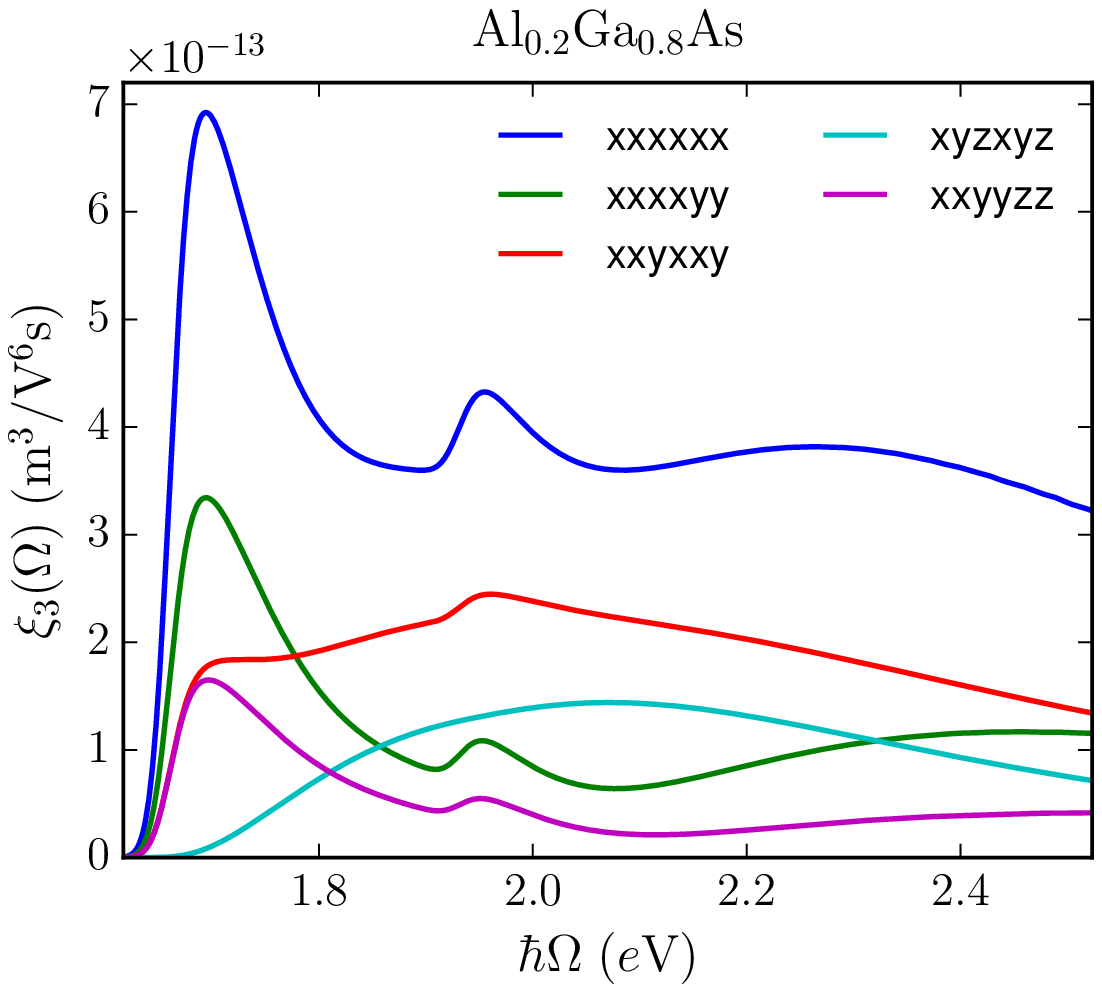}
\includegraphics[width=0.9\columnwidth]{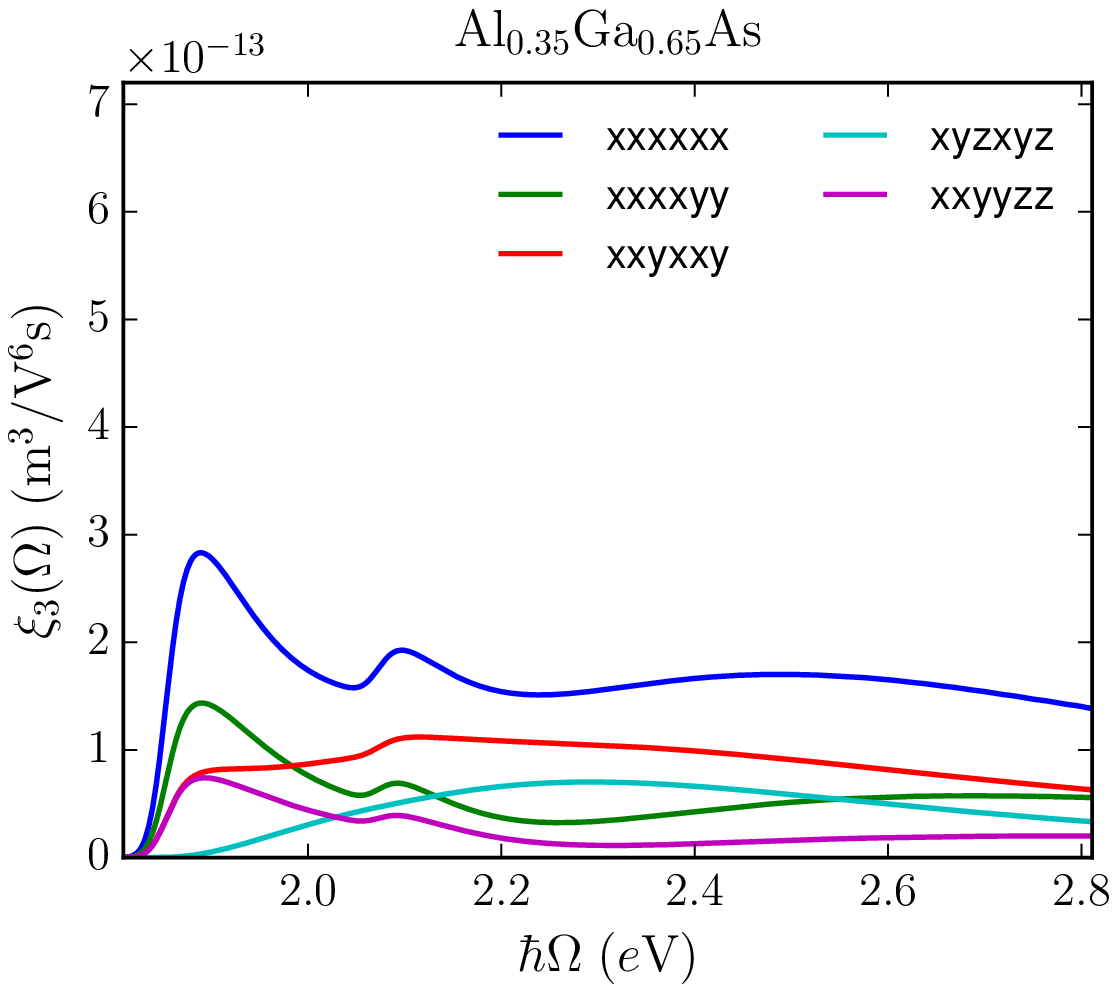}
\caption[]{ Three-photon carrier injection coefficients for two stoichiometries. }
\label{fig:carrier3} 
\end{figure}

\subsection{Carrier injection }

We track the number of injected carriers by calculating the number of electrons in the conduction bands, which corresponds to the operator  
\begin{equation}
\mathcal{N}=\sum_{c\bm{k}}a_{c\bm{k}}^{\dagger}a_{c\bm{k}},
\end{equation}
so we use $n_{cc^{\prime}}=\delta_{cc^{\prime}}$ and $n_{vv^{\prime}}=0$
for the carrier density matrix elements in Eq.~\eqref{eq:coeff}.
The optical injection of carriers due to 2PA and 3PA processes, as
well as their interference, is characterized by the tensors $\xi$,
according to 
\begin{align}
\frac{d}{dt}\left\langle n\right\rangle _{2}= & \xi_{2}^{abcd}\left(3\omega\right)E_{-3\omega/2}^{a}E_{-3\omega/2}^{b}E_{3\omega/2}^{c}E_{3\omega/2}^{d},\\
\frac{d}{dt}\left\langle n\right\rangle _{3}= & \xi_{3}^{abcdef}\left(3\omega\right)E_{-\omega}^{a}E_{-\omega}^{b}E_{-\omega}^{c}E_{\omega}^{d}E_{\omega}^{e}E_{\omega}^{f},\\
\frac{d}{dt}\left\langle n\right\rangle _{2+3(i)}= & \xi_{2+3}^{abcde}\left(3\omega\right)E_{-\omega}^{a}E_{-\omega}^{b}E_{-\omega}^{c}E_{3\omega/2}^{d}E_{3\omega/2}^{e} \nonumber \\ 
& +c.c.,
\end{align}
 where $\hbar\Omega=3\hbar\omega$ is the
total transition energy 
\footnote{If the band gap of the material is smaller than $5\omega/2$ there would be an additional contribution to $n_{2}$ proportional to $E_{-\omega}E_{-3\omega/2}E_{\omega}E_{3\omega/2}$, but these carriers do not contribute to the interference between 2PA and 3PA, which is our main interest. }, 
and the coefficients are calculated as 
\begin{align}
\xi_{2}^{abde}\left(3\omega\right)= & 2\pi\int\frac{d\bm{k}}{\left(2\pi\right)^{D}}\sum_{cv}R_{cv\bm{k}}^{ab\ast}R_{cv\bm{k}}^{de}\delta\left(3\omega-\omega_{cv}\right),\\
\xi_{3}^{abdefg}\left(3\omega\right)= & 2\pi\int\frac{d\bm{k}}{\left(2\pi\right)^{D}}\sum_{cv}R_{cv\bm{k}}^{abd\ast}R_{cv\bm{k}}^{efg}\delta\left(3\omega-\omega_{cv}\right),\\
\xi_{2+3}^{abdef}\left(3\omega\right)= & 2\pi\int\frac{d\bm{k}}{\left(2\pi\right)^{D}}\sum_{cv}R_{cv\bm{k}}^{abd\ast}R_{cv\bm{k}}^{ef}\delta\left(3\omega-\omega_{cv}\right).
\end{align}
The $\xi_{2+3}^{abdef}\left(3\omega\right)$ coefficient is associated with absorption processes described by $\chi^{(5)}$.  
The symmetries of the zincblende lattice, corresponding to the point group $T_{d}$ (or $43m$), strongly restrict the number of independent
non-zero components of the tensors $\xi_{2}$, $\xi_{2+3}$, and $\xi_{3}$.
We list the independent components of the injection tensor coefficients
in Appendix \ref{app:lattice}. In Figs.~\ref{fig:carrier2}, \ref{fig:carrier23}
and \ref{fig:carrier3}, we show the frequency dependence of the independent
components of the coefficients $\xi_{2}^{abcd}\left(3\omega\right)$, $\xi_{3}^{abcdef}\left(3\omega\right)$, and 
$\xi_{2+3}^{abcde}\left(3\omega\right)$  
respectively. Notice that the 3PA coefficient is large for frequencies
right above the band gap, while the coefficient for 2PA nearly vanishes
for similar frequencies. 
As discussed below Eqs.~\eqref{eq:R2coefSimp}
and \eqref{eq:R3coefSimp}, the dominant contribution to 2PA always
involves intraband velocity matrix elements, which correspond to the
band dispersion, so they vanish at the $\Gamma$ point of the Brillouin
zone. The 3PA has contributions from interband velocity matrix elements,
which in general do not vanish at $\Gamma$.

\begin{figure}[hbt!]
\includegraphics[width=0.9\columnwidth]{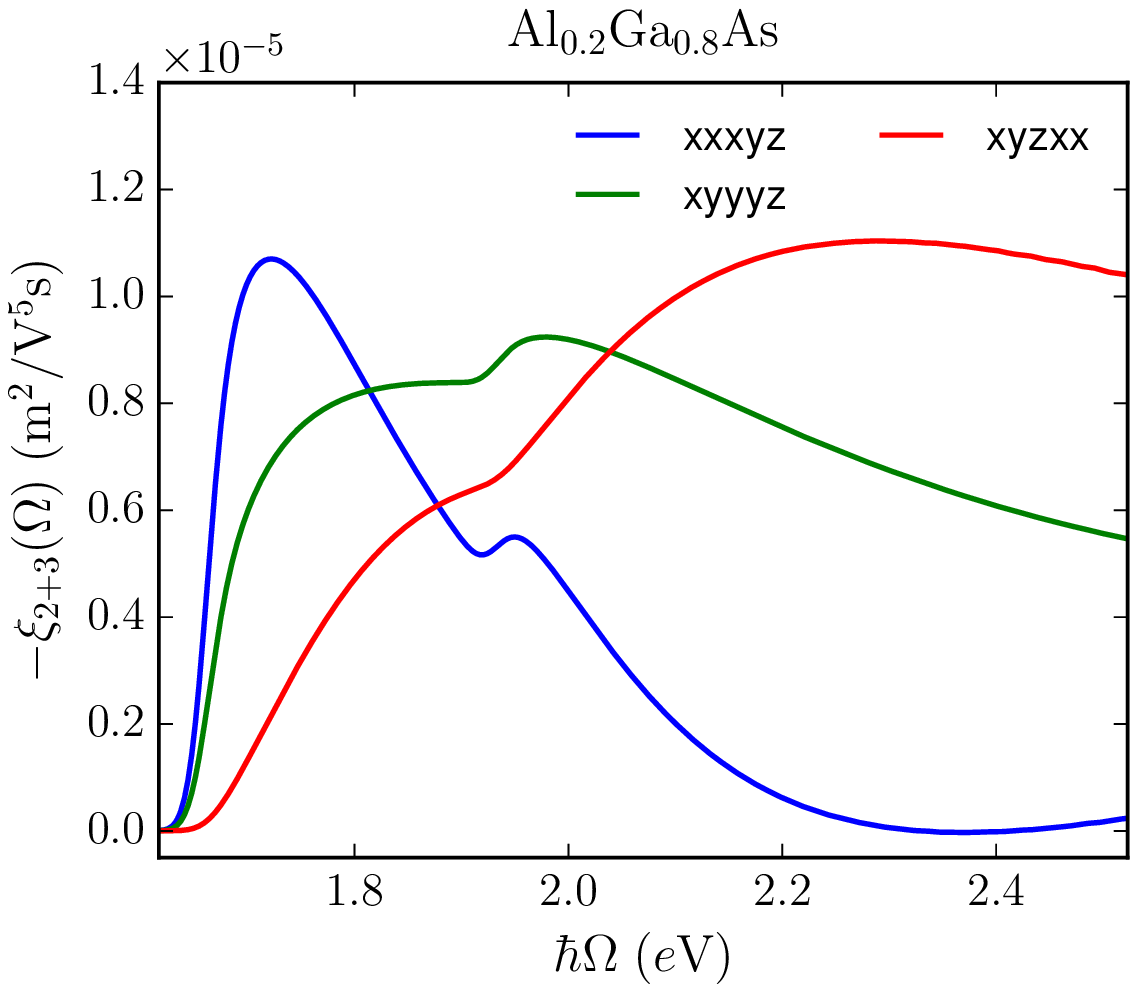}
\includegraphics[width=0.9\columnwidth]{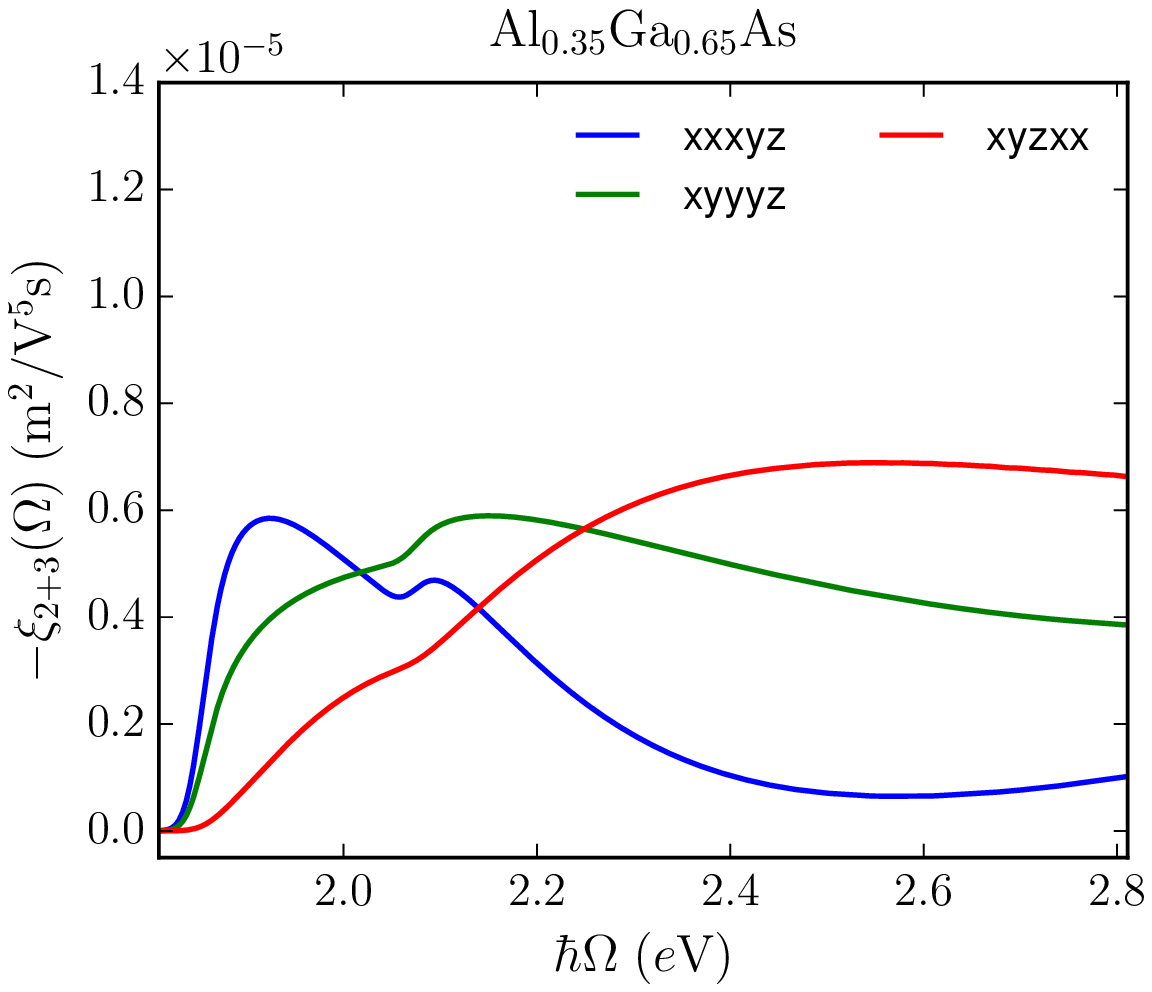}
\caption[]{ Injection rate coefficients for carrier density  corresponding to 2+3 QuIC for two stoichiometries. }
\label{fig:carrier23} 
\end{figure}

\begin{figure}[hbt!]
\includegraphics[width=0.9\columnwidth]{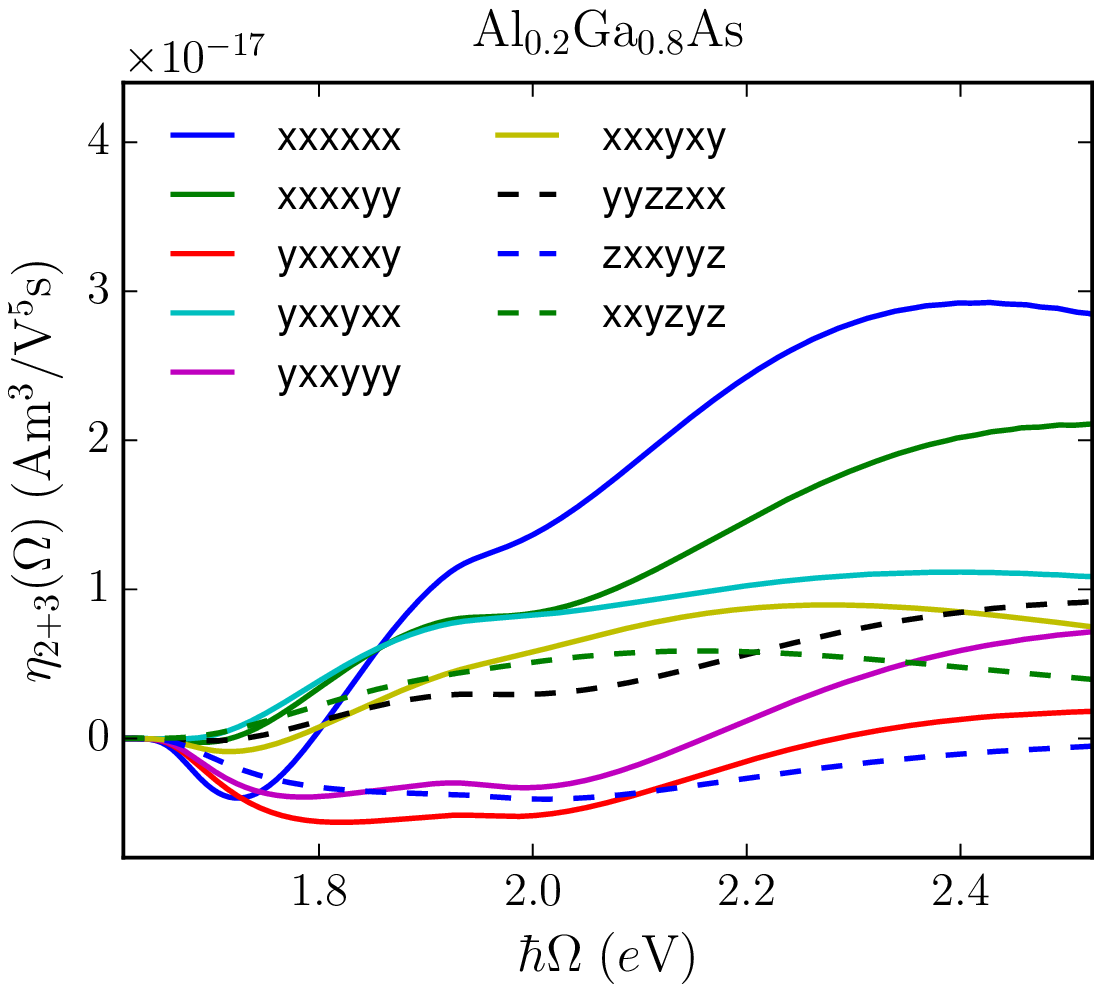}
\includegraphics[width=0.9\columnwidth]{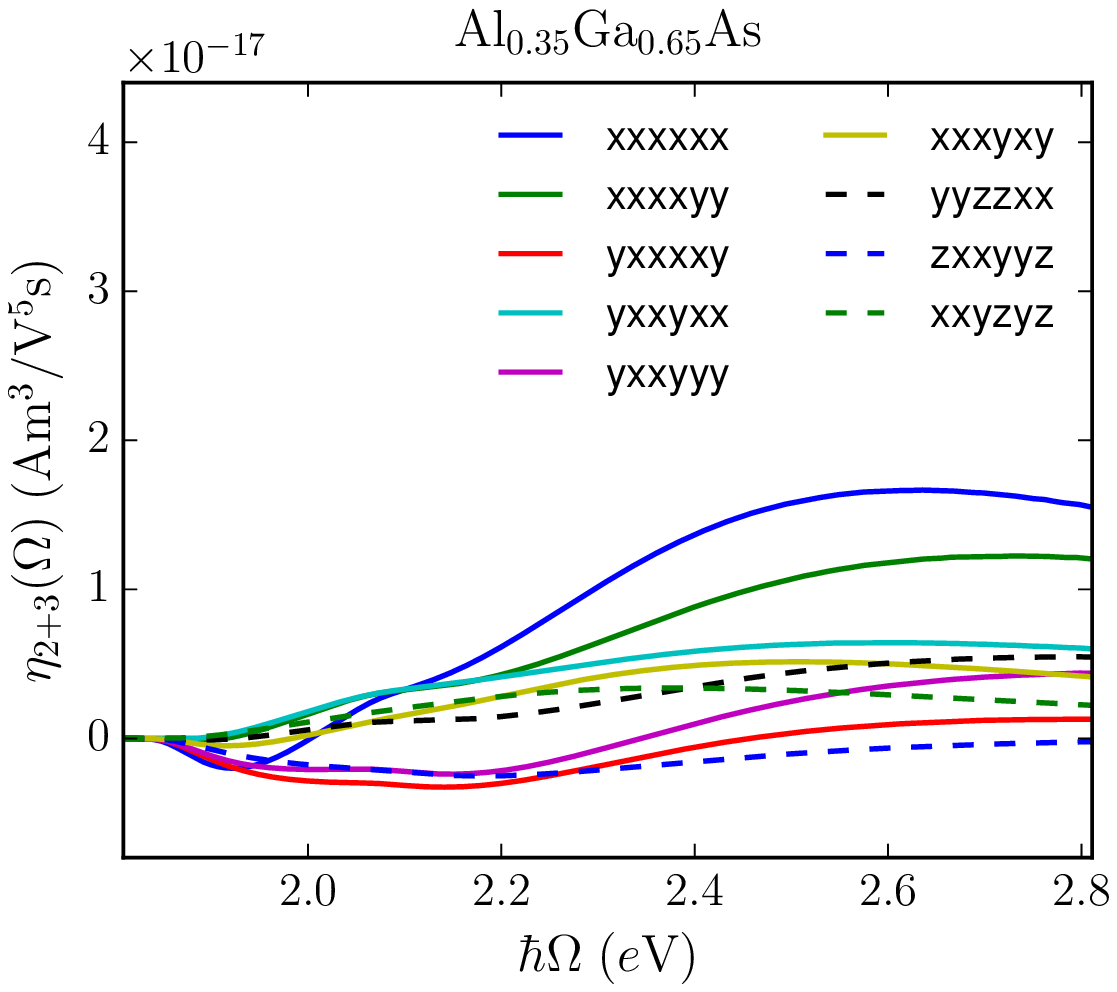}
\caption[]{ Injection rate coefficients for current  corresponding to 2+3 QuIC for two stoichiometries. }
\label{fig:current23} 
\end{figure}

\subsection{Current injection }

Taking the quantity $\left\langle M \right\rangle$ in Eq.~\eqref{eq:injectM} to be the current density $\left\langle \bm{J}\right\rangle $, we can compute its injection rate due to the quantum interference between two- and three-photon absorption processes as  
\begin{equation}
\frac{d}{dt}\left\langle J^{a}\right\rangle _{2+3}=\eta_{2+3}^{abcdef}\left(3\omega\right)E_{-\omega}^{b}E_{-\omega}^{c}E_{-\omega}^{d}E_{3\omega/2}^{e}E_{3\omega/2}^{f}+c.c.,
\label{eq:injectJ23}
\end{equation}
where $\hbar\Omega=3\hbar\omega$ is the total photon energy. 
We emphasize that $\eta_{2+3}^{abcdef}\left(3\omega\right)$ is related to $\chi^{(5)}$ and is finite even for centro-symmetric materials. 
The dependence of the injected current on the fields is described by Eq.~\eqref{eq:injectJ23}, which is the main equation to describe 2+3 QuIC experiments being reported in the accompanying article \cite{Wang17}.   
In terms of intensities, Eq.~\eqref{eq:injectJ23} indicates that the current injection rate is proportional to the intensity of the $3\omega/2$ field, $I_{3\omega/2}$, and the intensity of the $\omega$ field to the power $3/2$, $I_{\omega}^{3/2}$. These dependences are verified in Fig.~4 of the accompanying experimental article \cite{Wang17}, indicating that the currents measured in those experiments indeed correspond to 2+3 QuIC. 
We list the independent components of the injection tensor coefficient $\eta_{2+3}^{abcdef}\left(3\omega\right)$
in Appendix \ref{app:lattice}. In Fig.~\ref{fig:current23} we show
the frequency dependence of the independent components of the coefficient
$\eta_{2+3}^{abcdef}\left(3\omega\right)$ for different stoichiometries.
The plots show that some components change sign as the frequency increases.
This sign flip is due to the competing contributions due to intraband and interband
velocity matrix elements to the $R_{cv\bm{k}}^{\left(3\right)}$
coefficients. For low excess photon energies, the excited carriers
are close to the $\Gamma$ point in the BZ, and the interband contribution
is the most important, as the band dispersion is small. For larger
photon excess energies, the excited carriers are located further from
the $\Gamma$ point in the BZ, so the band dispersion is large and
the intraband contributions are more important.

\begin{figure}[hbtp!]
\includegraphics[width=0.9\columnwidth]{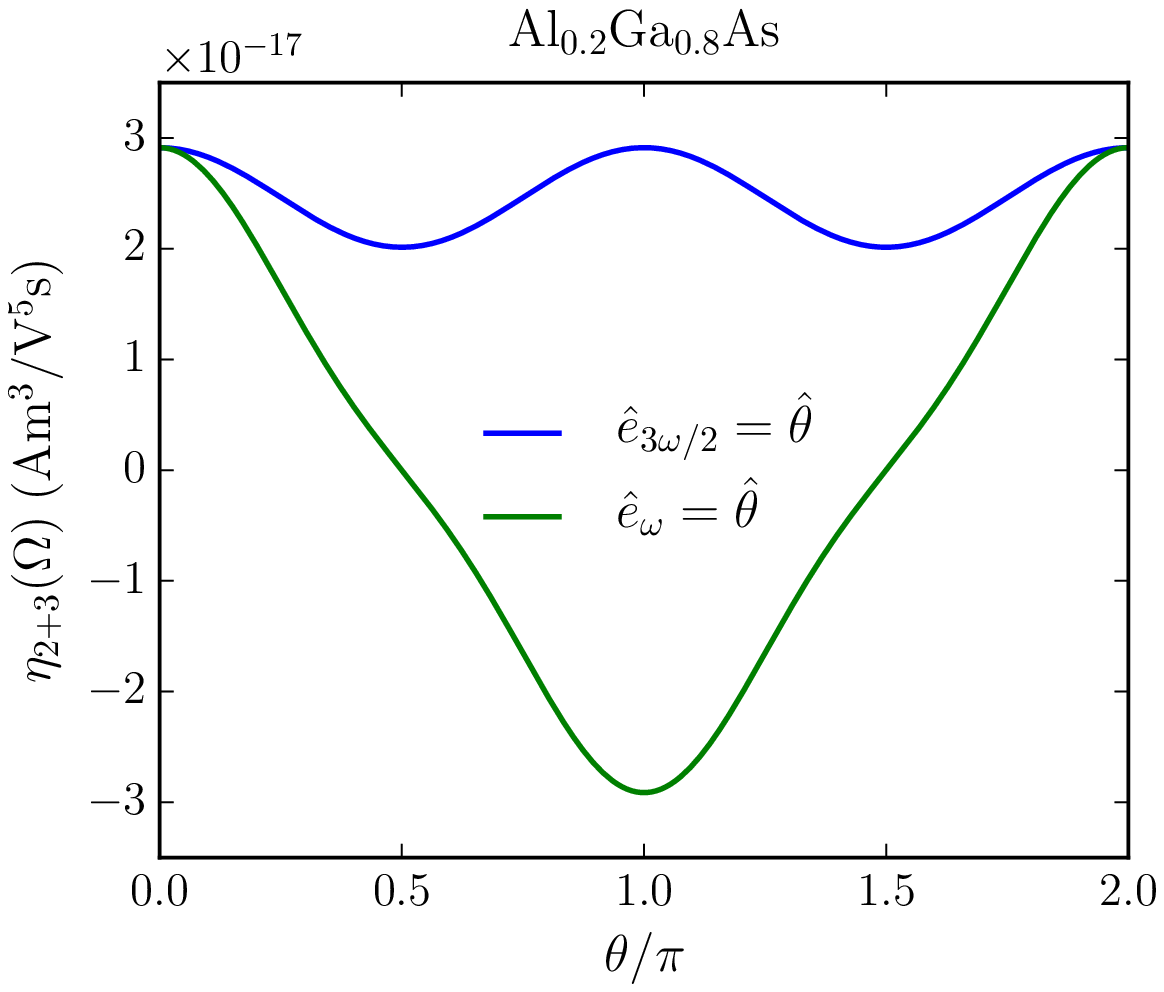}
\includegraphics[width=0.9\columnwidth]{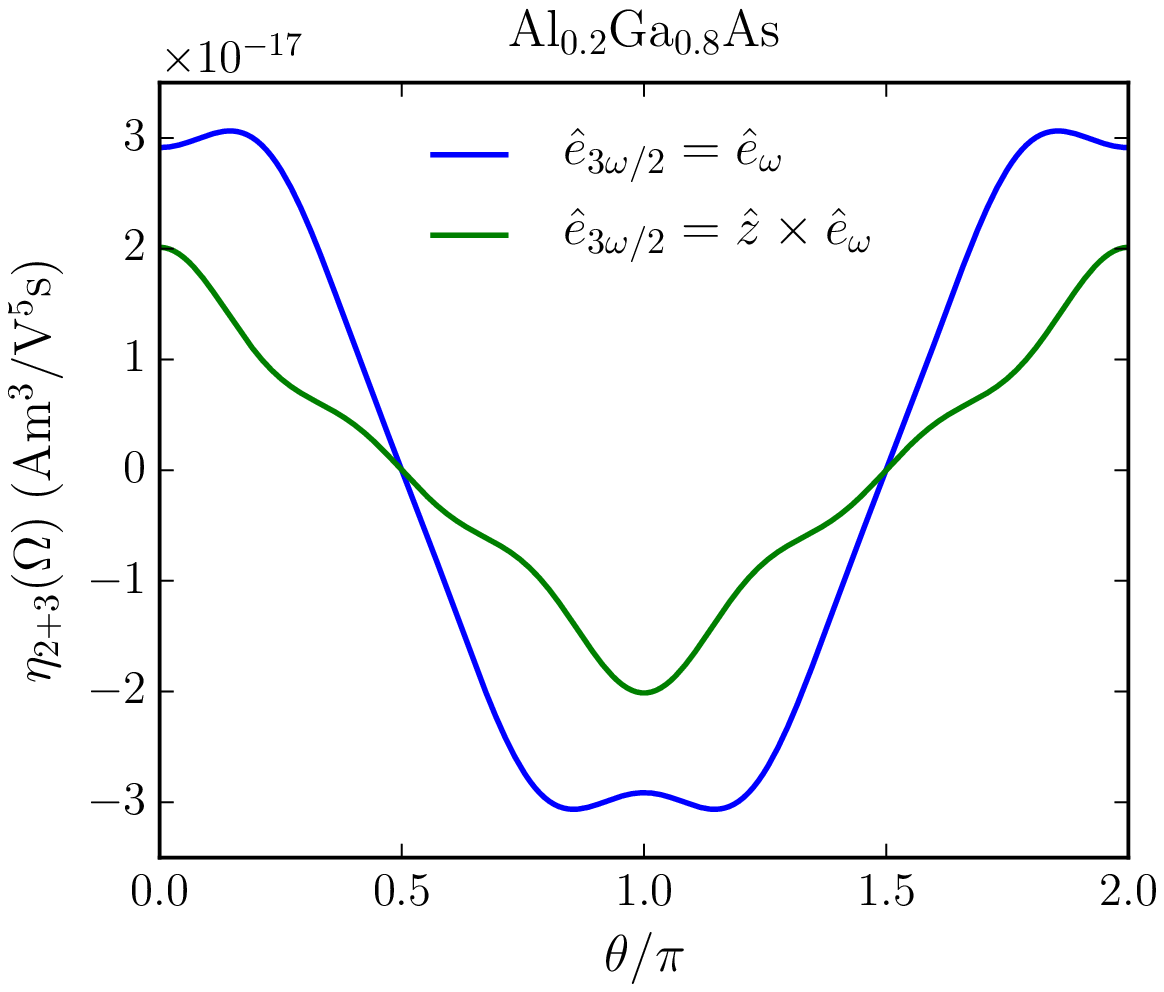}
\caption[]{ Injection rate coefficient for current along the $\hat{\bm{x}}$
direction as the polarizations of the incident fields are rotated
in the $\hat{\bm{x}}$-$\hat{\bm{y}}$ plane. (Top)
Either $\hat{\bm{e}}_{\omega}$ or $\hat{\bm{e}}_{3\omega/2}$
is rotated while the other is fixed along the $\hat{\bm{x}}$
direction. (Bottom) In both cases $\hat{\bm{e}}_{\omega}=\hat{\bm{\theta}}$
is rotated, and $\hat{\bm{e}}_{3\omega/2}$ is either parallel
or perpendicular to it. The total photon energy is $\hbar\Omega=2.4\,e$V
($\lambda \sim 520\,$nm) in both cases. }
\label{fig:polariz} 
\end{figure}

To illustrate some aspects of the different tensor components,
in Fig.~\ref{fig:polariz} we plot the injection current for different
polarizations of the incident fields in a typical experimental scenario.
We assume that the sample has electrodes mounted such that they always measure
the current along the [100] crystal direction, which we denote
by $\hat{\bm{x}}$. In the first case we keep either $\hat{\bm{e}}_{\omega}$
or $\hat{\bm{e}}_{3\omega/2}$ fixed along the $\hat{\bm{x}}$
direction, while the other field is rotated in the $\hat{\bm{x}}$-$\hat{\bm{y}}$
plane and points along the direction $\hat{\bm{\theta}}=\hat{\bm{x}}\cos\theta+\hat{\bm{y}}\sin\theta$,
where $\hat{\bm{y}}$ corresponds to the [010] crystal direction. 
The case where $\hat{\bm{e}}_{\omega} =\hat{\bm{\theta}}$
and $\hat{\bm{e}}_{3\omega/2} = \hat{\bm{x}}$, corresponding to the green line in Fig.~\ref{fig:polariz}, is tested experimentally and reported in Fig.~5 of the accompanying paper \cite{Wang17}. 
In the second scenario, the polarizations of both incident
fields are rotated in the $\hat{\bm{x}}$-$\hat{\bm{y}}$
plane and they are kept either parallel or perpendicular to each other.
In Fig.~\ref{fig:polariz}, we show that the current is largely along
the $\hat{\bm{e}}_{\omega}$ direction regardless of the $\hat{\bm{e}}_{3\omega/2}$
direction. However, the magnitude of the current depends significantly
on the $\hat{\bm{e}}_{3\omega/2}$ direction, and it is maximal
for $\hat{\bm{e}}_{3\omega/2}=\hat{\bm{e}}_{\omega}$.

\begin{figure}[hbtp!]
\includegraphics[width=0.9\columnwidth]{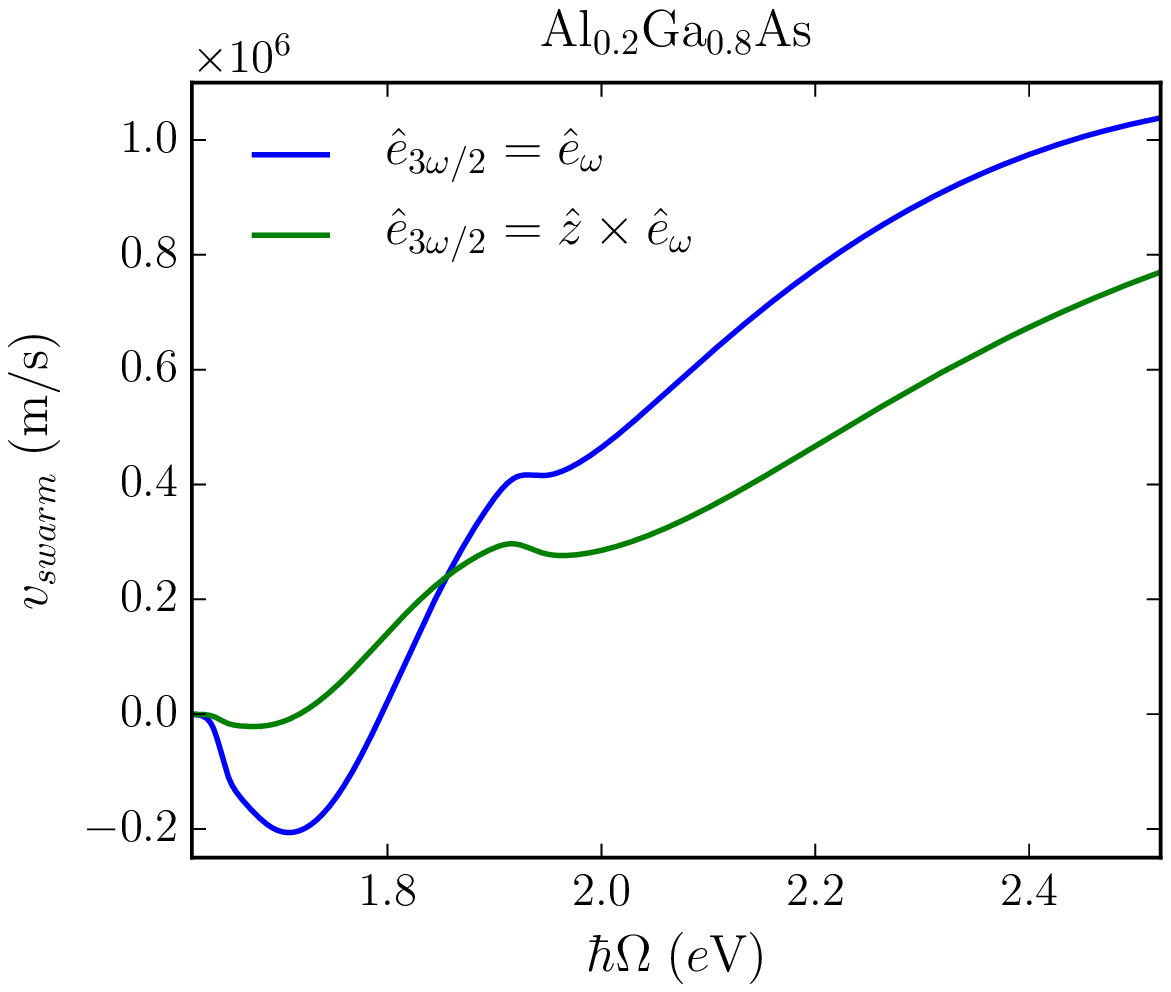}
\includegraphics[width=0.9\columnwidth]{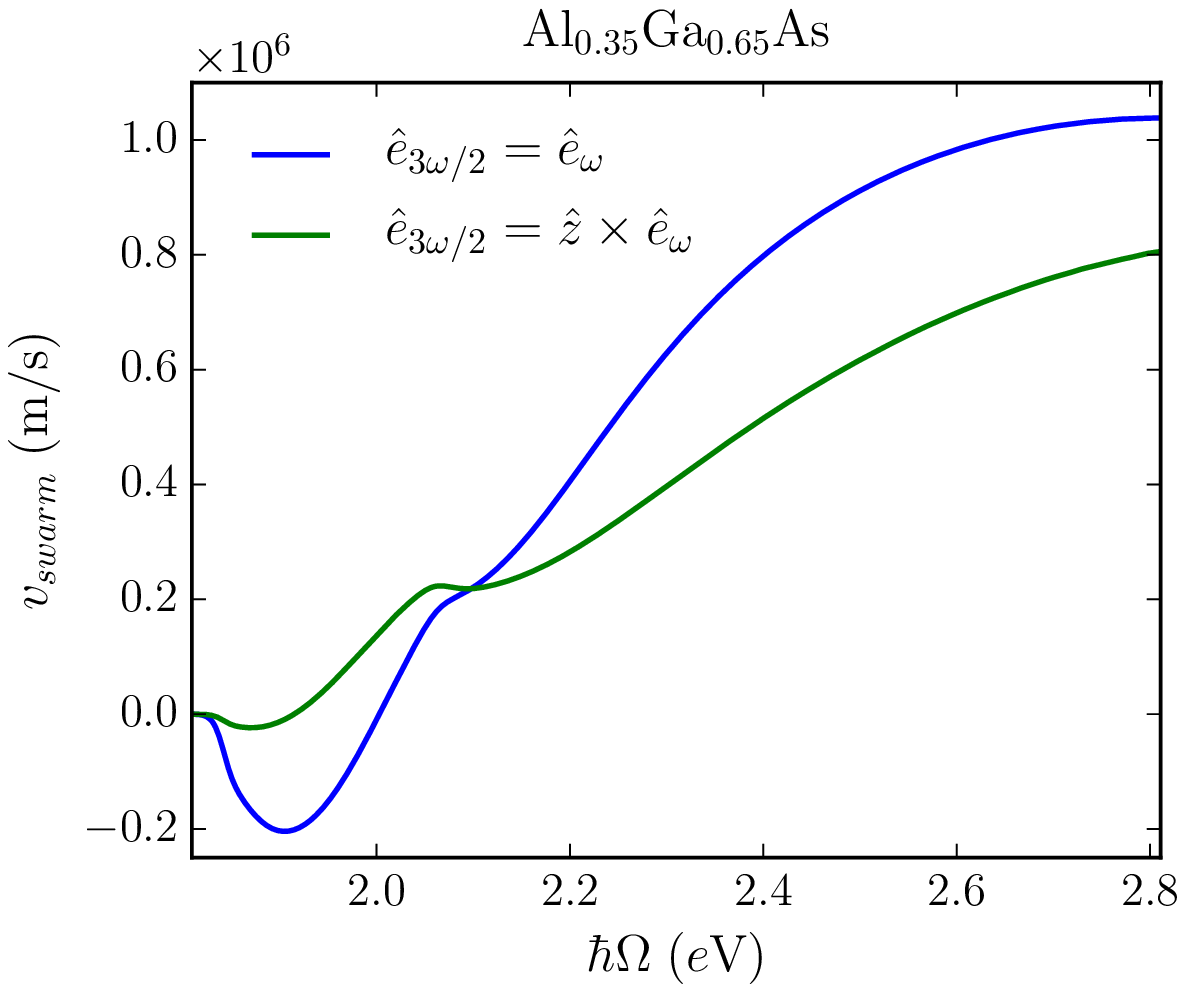}
\caption[]{ Swarm velocity for two stoichiometries assuming optimum interference between 2PA and 3PA.
The current is measured along the $\hat{\bm{x}}$ direction,
and so is the polarization of the lower frequency field $\hat{\bm{e}}_{\omega}=\hat{\bm{x}}$,
while we consider two cases for the polarization of the higher frequency
field: $\hat{\bm{e}}_{3\omega/2}=\hat{\bm{x}}$ and
$\hat{\bm{e}}_{3\omega/2}=\hat{\bm{y}}$. }
\label{fig:Vswarm23} 
\end{figure}

\subsubsection{Swarm velocity }

Since the excited carriers respond to the induced voltage due to the
injected current, and usually screen it at least partially, a good
measure of the efficiency of the current injection is the swarm velocity, defined as $\bm{v}_{{\rm swarm}}= \frac{d}{dt} \left\langle \bm{J}\right\rangle /e \frac{d}{dt} \left\langle n\right\rangle $, which represents the average contribution to the injection current due to one excited electron  
\footnote{ Notice that assuming constant injection rates, which is a reasonable assumption at least during some time interval, the definition of the swarm velocity is equivalent to 
 $ {\bf v}_{\rm swarm}=  <{\bf J}>  /e   <n>  $. }. 
Since $\left\langle n\right\rangle _{2+3}\ll\left\langle n\right\rangle _{2}+\left\langle n\right\rangle _{3}$,
the total density of carriers is $\left\langle n\right\rangle \simeq\left\langle n\right\rangle _{2}+\left\langle n\right\rangle _{3}$,
and for both light beams polarized along the $\bm{\hat{x}}$
direction we have a swarm velocity of magnitude 
\begin{equation}
v_{{\rm swarm}}=\dfrac{2\left|\eta_{2+3}^{xxxxxx}\left(3\omega\right)\right|E_{\omega}^{3}E_{3\omega/2}^{2}}{\left|e\right|\left(\xi_{3}^{xxxxxx}\left(3\omega\right)E_{\omega}^{6}+\xi_{2}^{xxxx}\left(3\omega\right)E_{3\omega/2}^{4}\right)},
\end{equation}
where we have chosen $\Delta\phi=\pi/2$ to optimize the magnitude
of the numerator. The whole expression is optimized by choosing the
intensities of the two beams appropriately; the condition to be satisfied
is $\xi_{2}^{xxxx}\left(3\omega\right)E_{3\omega/2}^{4}=\xi_{3}^{xxxxxx}\left(3\omega\right)E_{\omega}^{6}$
, which corresponds to an equal number of carriers injected by 2-photon
absorption and 3-photon absorption. If this holds, 
\begin{equation}
v_{{\rm swarm}}=\frac{\left|\eta_{2+3}^{xxxxxx}\left(3\omega\right)\right|}{\left|e\right|\sqrt{\xi_{3}^{xxxxxx}\left(3\omega\right)\xi_{2}^{xxxx}\left(3\omega\right)}}.\label{eq:swarm_all_x}
\end{equation}
In Fig.~\ref{fig:Vswarm23} we plot this expression, together with
the expression that would result if the beam of frequency $\omega$
were polarized in the \textbf{$\bm{\hat{x}}$} direction while
the one of frequency $3\omega/2$ in the $\bm{\hat{y}}$ direction,
which is the same as Eq.~\eqref{eq:swarm_all_x} but with $\eta_{2+3}^{xxxxxx}\left(3\omega\right)$
replaced by $\eta_{2+3}^{xxxxyy}\left(3\omega\right)$; as well, $\xi_{2}^{xxxx}\left(3\omega\right)$ should also be replaced
by $\xi_{2}^{yyyy}\left(3\omega\right)$, but they are equal. We see
that different stoichiometries give similar values for the swarm velocity
if the frequency is adjusted according to the band gap
of the system.  The fact that higher Ga concentrations lead to larger
injected currents (see Fig.~\ref{fig:current23}) is only due to
a higher carrier injection. Yet with appropriate laser intensities
it is possible to reach the same levels of injected current densities
with any Al concentration, although the laser frequencies and intensities
at which the maximum is achieved depend on the Al concentration.

We point out that the 2+3 QuIC swarm velocity is about twice its equivalent
for 1+2 QuIC. This is an indication that the distribution of carriers
injected in the BZ is sharper for 2+3 QuIC compared to 1+2 QuIC. We
further confirm that by computing the variance of the lattice momentum
$\bm{k}$ of the electrons injected in the conduction band for both
1+2 QuIC, $\sigma_{1+2}^{a}=\left\langle \left(k^{a}\right)^{2}\right\rangle _{1+2}-\left\langle k^{a}\right\rangle _{1+2}^{2}$,
and 2+3 QuIC, $\sigma_{2+3}^{a}=\left\langle \left(k^{a}\right)^{2}\right\rangle _{2+3}-\left\langle k^{a}\right\rangle _{2+3}^{2}$.
For the incident fields polarized along the $\hat{\bm{x}}$ direction,
we find $\left\langle \bm{k}\right\rangle _{1+2}=\left(4.9,\ 0,\ 0\right)\times10^{-2}{\rm \AA}^{-1}$
and $\left\langle \bm{k}\right\rangle _{2+3}=\left(5.7,\ 0,\ 0\right)\times10^{-2}{\rm \AA}^{-1}$,
as well as $\sigma_{1+2}=\left(3.4,\ 4.3,\ 4.3\right)\times10^{-3}{\rm \AA}^{-2}$
and $\sigma_{2+3}=\left(2.8,\ 2.2,\ 2.2\right)\times10^{-3}{\rm \AA}^{-2}$,
which indeed indicates that the distribution of injected electrons in
the BZ is sharper for 2+3 QuIC, especially in the directions transverse
($\hat{\bm{y}}$ and $\hat{\bm{z}}$) to the polarization of the field.

\subsubsection{Laser intensities }

Our calculations are performed in the perturbative regime, the validity
of which requires that the fraction of the injected carrier population
density relative to the total density of states $n_{{\rm max}}$ in the
range of energies covered by the laser pulse be small. We thus consider
our calculations to be valid when 
\begin{equation}
\left\langle n\right\rangle _{2}+\left\langle n\right\rangle _{3}<0.1n_{{\rm max}},\label{eq:maxDens}
\end{equation}
where the fraction $0.1$ is chosen somewhat arbitrarily. The carrier
injection due to the 2- and 3-photon interference $\left\langle n\right\rangle _{2+3 (i)}$
mostly has the effect of concentrating the carrier injection in some
region of the BZ, but it does not contribute significantly to the
total number of injected carriers compared to $\left\langle n\right\rangle _{2}$
and $\left\langle n\right\rangle _{3}$. For the estimates of laser
intensities we consider the incident fields to be both polarized along
the $\hat{\bm{x}}$ direction, so for a laser pulse of duration ${\cal T}$ 
we require 
\begin{align}
\left[\frac{d}{dt}\left\langle n\right\rangle _{2}+\frac{d}{dt}\left\langle n\right\rangle _{3}\right]{\cal T}< & 0.1n_{{\rm max}},\\
\left[\xi_{2}^{xxxx}\left(3\omega\right)E_{3\omega/2}^{4}+\xi_{3}^{xxxxxx}\left(3\omega\right)E_{\omega}^{6}\right]{\cal T}< & 0.1n_{{\rm max}}.\label{eq:maxRate}
\end{align}
The maximum density of states $n_{{\rm max}}$ that can be injected is determined
by analyzing the volume $\mathtt{V}$ corresponding to the excited
states in the BZ. We denote by $k_{\Omega}$ the momentum corresponding
to the energy difference $\hbar\omega_{cv\bm{k}}=\hbar\Omega$ between
the conduction and valence bands, so $\mathtt{V}=4\pi k_{\Omega}^{2}\Delta k$,
where $\Delta k=\frac{dk}{d\omega_{cv\bm{k}}}\Delta\omega_{cv\bm{k}}$
is related to the frequency broadening $\Delta\omega=2\pi/{\cal T}$
associated with the time duration of the pulse. The derivative of
the band energy corresponds to the velocities of electrons in the
conduction and valence bands, $v_{\Omega}=\frac{d\omega_{cv}}{dk}=\frac{d\omega_{c}}{dk}-\frac{d\omega_{v}}{dk}$,
so $\mathtt{V}=8\pi^{2}k_{\Omega}^{2}/v_{\Omega}{\cal T}$. The volume
in the BZ associated with one quantum state is $\mathtt{V}_{1}=\left(2\pi/L\right)^{3}$,
where $L$ is the normalization length of the sample. The number of
states that can be excited is then $\mathtt{V}/\mathtt{V}_{1}$,
and their spatial density is 
\begin{equation}
n_{max}=\frac{\mathtt{V}}{\mathtt{V}_{1}L^{3}}=\frac{k_{\Omega}^{2}}{\pi v_{\Omega}{\cal T}}.
\end{equation}
For optimal interference, there should be equal densities of carriers
injected by 2- and 3-photon absorption, $\left\langle n\right\rangle _{2}=\left\langle n\right\rangle _{3}$,
which according to Eq.~\eqref{eq:maxRate} gives 
\begin{equation}
\xi_{2}^{xxxx}\left(3\omega\right)E_{3\omega/2}^{4}=\xi_{3}^{xxxxxx}\left(3\omega\right)E_{\omega}^{6}<0.05\frac{k_{\Omega}^{2}}{\pi v_{\Omega}{\cal T}^{2}}.
\end{equation}
The maximal amplitudes $E_{\omega}$ and $E_{3\omega/2}$ of the incident
fields can then be estimated from the extreme of the inequality in
the above equation. For the stoichiometry of $\alpha=0.2$, pulses
with duration ${\cal T}=150\,$fs, and total photon energy $\hbar\Omega=2.4\,e$V,
we have 
\begin{align}
E_{\omega}= & 1.24\times10^{8}\,\frac{\rm V}{\rm m},\\
E_{3\omega/2}= & 6.05\times10^{7}\,\frac{\rm V}{\rm m}.
\end{align}
The intensities in the material medium with these field amplitudes
are 
\begin{align}
I_{\omega}= & 2\epsilon_{0}cn_{\omega}E_{\omega}^{2}=26.5\,\frac{{\rm GW}}{{\rm cm}^{2}},\\
I_{3\omega/2}= & 2\epsilon_{0}cn_{3\omega/2}E_{3\omega/2}^{2}=6.54\,\frac{{\rm GW}}{{\rm cm}^{2}}.
\end{align}
For these values, the injected current density is 
\begin{equation}
\left\langle J^{x}\right\rangle =2\eta_{2+3}^{xxxxxx}\left(3\omega\right)E_{\omega}^{3}E_{3\omega/2}^{2}{\cal T}=6.25\,\frac{{\rm MA}}{{\rm cm}^{2}}.
\end{equation}
We emphasize that these are just estimates, as the limit of carrier
density is set somewhat arbitrarily in Eq.~\eqref{eq:maxDens}. We
note that we are ignoring scattering of the injected carriers. This
means that the true maximal intensities would be larger than our estimates
here, since there is room for more photon absorption as scattering
depletes some of the excited states. We also note that in this treatment
the electron-electron interaction has been neglected; were it included,
the phase parameter would be shifted. However this shift is usually
very small for zincblende semiconductors, except for frequencies very
close to the band gap \cite{bhat05}.

\section{Discussion and conclusion }

\label{sec:conclusion}

One of the main utilities of QuIC in semiconductors is the injection of carriers in localized regions of the BZ. In this respect 2+3 QuIC performs better than 1+2 QuIC, and that can be seen in a higher swarm velocity of 2+3 QuIC, which is a desirable feature for current injection. 
Another interesting difference between 1+2 and 2+3 QuIC is that in 2+3 QuIC several current injection coefficients change sign as the total photon energy
is increased, while in 1+2 QuIC they typically do not. 
This happens because interband velocity matrix elements are responsible for the largest contribution to the 3PA coefficient at low photon energies, but at higher photon energies the intraband velocity matrix elements dominate. 
Since only nonlinear optical processes are involved in 2+3 QuIC, the laser intensities
required for maximal effect are higher than for 1+2 QuIC, but still moderate. 
Also, the optical fields have a power law attenuation as they propagate through the absorbing material, instead of the exponential attenuation
of linear absorption. 
Thus a waveguide geometry is desirable, and while QuIC in waveguides presents some challenges, as it raises issues of phase- and mode-matching, it also presents opportunities for easy integration with devices on-chip. 
Since optical frequency combs are routinely propagated through waveguides, there should be no additional
difficulties for 2+3 QuIC experiments in waveguides other than the usual issues of phase- and mode-matching.

\section*{Acknowledgment}

We thank Perry T. Mahon for useful discussions. 
This work has been supported by DARPA through the DODOS program.

\appendix

\section{Nonzero injection coefficient components of zincblende lattices }

\label{app:lattice}

AlGaAs in the virtual crystal approximation forms a zincblende lattice,
which has the symmetry of point group $T_{d}$ (or $43m$). The optical
responses we consider in this work involve tensors of rank 4 up to
6. With $T_{d}$ symmetries \cite{Sauter96}, generic rank-4 tensors
have 21 non-zero components of which 4 are independent, rank-5 tensors
have 60 non-zero and 10 independent components, and rank-6 tensors
have 183 non-zero and 31 independent components. However, the tensors
representing the optical processes have a few more specific restrictions
due to their relation to the optical fields, as the indices associated
with the same incident field are symmetrized.

With these considerations the tensor $\xi_{2}$ has 3 independent
components 
\begin{align}
\xi_{2}^{xxxx}= & P\left(x,y,z\right),\\
\xi_{2}^{xyxy}= & \xi_{2}^{xyyx}=P\left(x,y,z\right),\\
\xi_{2}^{xxyy}= & P\left(x,y,z\right),
\end{align}
where $P\left(x,y,z\right)$ indicates all the possible permutations
of $\left(x,y,z\right)$ in the indices. The tensor $\xi_{2+3}$ has
3 independent components 
\begin{align}
\xi_{2+3}^{xxxyz}= & P\left(x,y,z\right),\\
\xi_{2+3}^{xxyxz}= & \xi_{2+3}^{xxyzx}=\xi_{2+3}^{xyxxz}\nonumber \\
= & \xi_{2+3}^{xyxzx}=\xi_{2+3}^{yxxxz}=\xi_{2+3}^{yxxzx}=P\left(x,y,z\right),\\
\xi_{2+3}^{xyzxx}= & \xi_{2+3}^{yxzxx}=\xi_{2+3}^{yzxxx}=P\left(x,y,z\right),
\end{align}
and $\xi_{3}$ has 5 independent components 
\begin{align}
\xi_{3}^{xxxxxx}= & P\left(x,y,z\right),\\
\xi_{3}^{xxxxyy}= & \xi_{3}^{xxxyxy}=\xi_{3}^{xxxyyx}\nonumber \\
= & \xi_{3}^{yyxxxx}=\xi_{3}^{yxyxxx}=\xi_{3}^{xyyxxx}=P\left(x,y,z\right),\\
\xi_{3}^{xxyxxy}= & \xi_{3}^{xxyxyx}=\xi_{3}^{xxyyxx}=\xi_{3}^{xyxxxy}=\xi_{3}^{xyxxyx}=\xi_{3}^{xyxyxx}\nonumber \\
= & \xi_{3}^{yxxxxy}=\xi_{3}^{yxxxyx}=\xi_{3}^{yxxyxx}=P\left(x,y,z\right),\\
\xi_{3}^{xxyyzz}= & \xi_{3}^{xxyzyz}=\xi_{3}^{xxyzzy}=\xi_{3}^{xyxyzz}=\xi_{3}^{xyxyzyz}=\xi_{3}^{xyxzzy}\nonumber \\
= & \xi_{3}^{yxxyzz}=\xi_{3}^{yxxyzyz}=\xi_{3}^{yxxzzy}=P\left(x,y,z\right),\\
\xi_{3}^{xyzxyz}= & \xi_{3}^{xyzzxy}=\xi_{3}^{xyzyzx}=\xi_{3}^{xyzzyx}\nonumber \\
= & \xi_{3}^{xyzxzy}=\xi_{3}^{xyzyxz}=P\left(x,y,z\right).
\end{align}
Finally, the tensor $\eta_{2+3}$ has 9 independent components 
\begin{align}
\eta_{2+3}^{xxxxxx}= & P\left(x,y,z\right),\\
\eta_{2+3}^{xxxxyy}= & P\left(x,y,z\right),\\
\eta_{2+3}^{xxyyxx}= & \eta_{2+3}^{xyxyxx}=\eta_{2+3}^{xyyxxx}=P\left(x,y,z\right),\\
\eta_{2+3}^{xxxyxy}= & \eta_{2+3}^{xxxyyx}=\eta_{2+3}^{xxyxxy}\nonumber \\
= & \eta_{2+3}^{xxyxyx}=\eta_{2+3}^{xyxxxy}=\eta_{2+3}^{xyxxyx}=P\left(x,y,z\right),\\
\eta_{2+3}^{yxxxxy}= & \eta_{2+3}^{yxxxyx}=P\left(x,y,z\right),\\
\eta_{2+3}^{yxxyxx}= & \eta_{2+3}^{yxyxxx}=\eta_{2+3}^{yyxxxx}=P\left(x,y,z\right),\\
\eta_{2+3}^{xxyyzz}= & \eta_{2+3}^{xyxyzz}=\eta_{2+3}^{xyyxzz}=P\left(x,y,z\right),\\
\eta_{2+3}^{xxyzyz}= & \eta_{2+3}^{xxzyyz}=\eta_{2+3}^{xyxzyz}\nonumber \\
= & \eta_{2+3}^{xyzxyz}=\eta_{2+3}^{xzxyyz}=\eta_{2+3}^{xzyxyz}=P\left(x,y,z\right),\\
\eta_{2+3}^{xyyzxz}= & \eta_{2+3}^{xyzyxz}=\eta_{2+3}^{xzyyxz}\nonumber \\
= & \eta_{2+3}^{xyyzzx}=\eta_{2+3}^{xyzyzx}=\eta_{2+3}^{xzyyzx}=P\left(x,y,z\right).
\end{align}
The independent components shown here are the ones plotted in the
figures in the main text.

\bibliography{bibliQuIC}

\begin{thebibliography}{52}%
\makeatletter
\providecommand \@ifxundefined [1]{%
 \@ifx{#1\undefined}
}%
\providecommand \@ifnum [1]{%
 \ifnum #1\expandafter \@firstoftwo
 \else \expandafter \@secondoftwo
 \fi
}%
\providecommand \@ifx [1]{%
 \ifx #1\expandafter \@firstoftwo
 \else \expandafter \@secondoftwo
 \fi
}%
\providecommand \natexlab [1]{#1}%
\providecommand \enquote  [1]{``#1''}%
\providecommand \bibnamefont  [1]{#1}%
\providecommand \bibfnamefont [1]{#1}%
\providecommand \citenamefont [1]{#1}%
\providecommand \href@noop [0]{\@secondoftwo}%
\providecommand \href [0]{\begingroup \@sanitize@url \@href}%
\providecommand \@href[1]{\@@startlink{#1}\@@href}%
\providecommand \@@href[1]{\endgroup#1\@@endlink}%
\providecommand \@sanitize@url [0]{\catcode `\\12\catcode `\$12\catcode
  `\&12\catcode `\#12\catcode `\^12\catcode `\_12\catcode `\%12\relax}%
\providecommand \@@startlink[1]{}%
\providecommand \@@endlink[0]{}%
\providecommand \url  [0]{\begingroup\@sanitize@url \@url }%
\providecommand \@url [1]{\endgroup\@href {#1}{\urlprefix }}%
\providecommand \urlprefix  [0]{URL }%
\providecommand \Eprint [0]{\href }%
\providecommand \doibase [0]{http://dx.doi.org/}%
\providecommand \selectlanguage [0]{\@gobble}%
\providecommand \bibinfo  [0]{\@secondoftwo}%
\providecommand \bibfield  [0]{\@secondoftwo}%
\providecommand \translation [1]{[#1]}%
\providecommand \BibitemOpen [0]{}%
\providecommand \bibitemStop [0]{}%
\providecommand \bibitemNoStop [0]{.\EOS\space}%
\providecommand \EOS [0]{\spacefactor3000\relax}%
\providecommand \BibitemShut  [1]{\csname bibitem#1\endcsname}%
\let\auto@bib@innerbib\@empty
\bibitem [{\citenamefont {Manykin}(2001)}]{Manykin}%
  \BibitemOpen
  \bibfield  {author} {\bibinfo {author} {\bibfnamefont {E.~A.}\ \bibnamefont
  {Manykin}},\ }\href@noop {} {\bibfield  {journal} {\bibinfo  {journal} {Laser
  Physics}\ }\textbf {\bibinfo {volume} {11}},\ \bibinfo {pages} {60} (\bibinfo
  {year} {2001})}\BibitemShut {NoStop}%
\bibitem [{\citenamefont {Shapiro}\ and\ \citenamefont
  {Brumer}(2003)}]{Shapiro03}%
  \BibitemOpen
  \bibfield  {author} {\bibinfo {author} {\bibfnamefont {M.}~\bibnamefont
  {Shapiro}}\ and\ \bibinfo {author} {\bibfnamefont {P.}~\bibnamefont
  {Brumer}},\ }\href@noop {} {\emph {\bibinfo {title} {Principles of the
  Quantum Control of Molecular Processes}}}\ (\bibinfo  {publisher}
  {Wiley-Interscience, New York},\ \bibinfo {year} {2003})\BibitemShut
  {NoStop}%
\bibitem [{\citenamefont {Brumer}\ and\ \citenamefont
  {Shapiro}(1986)}]{Brumer86}%
  \BibitemOpen
  \bibfield  {author} {\bibinfo {author} {\bibfnamefont {P.}~\bibnamefont
  {Brumer}}\ and\ \bibinfo {author} {\bibfnamefont {M.}~\bibnamefont
  {Shapiro}},\ }\href@noop {} {\bibfield  {journal} {\bibinfo  {journal}
  {Chemical Physics Letters}\ }\textbf {\bibinfo {volume} {126}},\ \bibinfo
  {pages} {541} (\bibinfo {year} {1986})}\BibitemShut {NoStop}%
\bibitem [{\citenamefont {Chen}\ \emph {et~al.}(1990)\citenamefont {Chen},
  \citenamefont {Yin},\ and\ \citenamefont {Elliott}}]{Chen90}%
  \BibitemOpen
  \bibfield  {author} {\bibinfo {author} {\bibfnamefont {C.}~\bibnamefont
  {Chen}}, \bibinfo {author} {\bibfnamefont {Y.-Y.}\ \bibnamefont {Yin}}, \
  and\ \bibinfo {author} {\bibfnamefont {D.~S.}\ \bibnamefont {Elliott}},\
  }\href@noop {} {\bibfield  {journal} {\bibinfo  {journal} {Physical Review
  Letters}\ }\textbf {\bibinfo {volume} {64}},\ \bibinfo {pages} {507}
  (\bibinfo {year} {1990})}\BibitemShut {NoStop}%
\bibitem [{\citenamefont {Zhu}\ \emph {et~al.}(1995)\citenamefont {Zhu},
  \citenamefont {Kleiman}, \citenamefont {Li}, \citenamefont {Lu},
  \citenamefont {Trentelman},\ and\ \citenamefont {Gordon}}]{Zhu95}%
  \BibitemOpen
  \bibfield  {author} {\bibinfo {author} {\bibfnamefont {L.}~\bibnamefont
  {Zhu}}, \bibinfo {author} {\bibfnamefont {V.}~\bibnamefont {Kleiman}},
  \bibinfo {author} {\bibfnamefont {X.}~\bibnamefont {Li}}, \bibinfo {author}
  {\bibfnamefont {S.~P.}\ \bibnamefont {Lu}}, \bibinfo {author} {\bibfnamefont
  {K.}~\bibnamefont {Trentelman}}, \ and\ \bibinfo {author} {\bibfnamefont
  {R.~J.}\ \bibnamefont {Gordon}},\ }\href@noop {} {\bibfield  {journal}
  {\bibinfo  {journal} {Science}\ }\textbf {\bibinfo {volume} {270}},\ \bibinfo
  {pages} {77} (\bibinfo {year} {1995})}\BibitemShut {NoStop}%
\bibitem [{\citenamefont {Nagai}\ \emph {et~al.}(2006)\citenamefont {Nagai},
  \citenamefont {Ohmura}, \citenamefont {Ito}, \citenamefont {Nakanaga},\ and\
  \citenamefont {Tachiya}}]{Nagai06}%
  \BibitemOpen
  \bibfield  {author} {\bibinfo {author} {\bibfnamefont {H.}~\bibnamefont
  {Nagai}}, \bibinfo {author} {\bibfnamefont {H.}~\bibnamefont {Ohmura}},
  \bibinfo {author} {\bibfnamefont {F.}~\bibnamefont {Ito}}, \bibinfo {author}
  {\bibfnamefont {T.}~\bibnamefont {Nakanaga}}, \ and\ \bibinfo {author}
  {\bibfnamefont {M.}~\bibnamefont {Tachiya}},\ }\href@noop {} {\bibfield
  {journal} {\bibinfo  {journal} {J. Chem. Phys.}\ }\textbf {\bibinfo {volume}
  {124}},\ \bibinfo {pages} {034304} (\bibinfo {year} {2006})}\BibitemShut
  {NoStop}%
\bibitem [{\citenamefont {Kurizki}\ \emph {et~al.}(1989)\citenamefont
  {Kurizki}, \citenamefont {Shapiro},\ and\ \citenamefont
  {Brumer}}]{Kurizki89}%
  \BibitemOpen
  \bibfield  {author} {\bibinfo {author} {\bibfnamefont {G.}~\bibnamefont
  {Kurizki}}, \bibinfo {author} {\bibfnamefont {M.}~\bibnamefont {Shapiro}}, \
  and\ \bibinfo {author} {\bibfnamefont {P.}~\bibnamefont {Brumer}},\
  }\href@noop {} {\bibfield  {journal} {\bibinfo  {journal} {Physical Review
  B}\ }\textbf {\bibinfo {volume} {39}},\ \bibinfo {pages} {3435} (\bibinfo
  {year} {1989})}\BibitemShut {NoStop}%
\bibitem [{\citenamefont {Baranova}\ \emph {et~al.}(1990)\citenamefont
  {Baranova}, \citenamefont {Chudinov},\ and\ \citenamefont
  {Zel'dovich}}]{Baranova90}%
  \BibitemOpen
  \bibfield  {author} {\bibinfo {author} {\bibfnamefont {N.~B.}\ \bibnamefont
  {Baranova}}, \bibinfo {author} {\bibfnamefont {A.~N.}\ \bibnamefont
  {Chudinov}}, \ and\ \bibinfo {author} {\bibfnamefont {B.~Y.}\ \bibnamefont
  {Zel'dovich}},\ }\href@noop {} {\bibfield  {journal} {\bibinfo  {journal}
  {Optics Communications}\ }\textbf {\bibinfo {volume} {79}},\ \bibinfo {pages}
  {116} (\bibinfo {year} {1990})}\BibitemShut {NoStop}%
\bibitem [{\citenamefont {Lawandy}(1991)}]{Lawandy90}%
  \BibitemOpen
  \bibfield  {author} {\bibinfo {author} {\bibfnamefont {N.~M.}\ \bibnamefont
  {Lawandy}},\ }\href@noop {} {\bibfield  {journal} {\bibinfo  {journal}
  {Optics Communications}\ }\textbf {\bibinfo {volume} {85}},\ \bibinfo {pages}
  {369} (\bibinfo {year} {1991})}\BibitemShut {NoStop}%
\bibitem [{\citenamefont {Baranova}\ \emph {et~al.}(1991)\citenamefont
  {Baranova}, \citenamefont {Chudinov},\ and\ \citenamefont
  {Zel'dovich}}]{Baranova91}%
  \BibitemOpen
  \bibfield  {author} {\bibinfo {author} {\bibfnamefont {N.~B.}\ \bibnamefont
  {Baranova}}, \bibinfo {author} {\bibfnamefont {A.~N.}\ \bibnamefont
  {Chudinov}}, \ and\ \bibinfo {author} {\bibfnamefont {B.~Y.}\ \bibnamefont
  {Zel'dovich}},\ }\href@noop {} {\bibfield  {journal} {\bibinfo  {journal}
  {Optics Communications}\ }\textbf {\bibinfo {volume} {85}},\ \bibinfo {pages}
  {371} (\bibinfo {year} {1991})}\BibitemShut {NoStop}%
\bibitem [{\citenamefont {Dupont}\ \emph {et~al.}(1995)\citenamefont {Dupont},
  \citenamefont {Corkum}, \citenamefont {Liu}, \citenamefont {Buchanan},\ and\
  \citenamefont {Wasilewski}}]{Dupont95}%
  \BibitemOpen
  \bibfield  {author} {\bibinfo {author} {\bibfnamefont {E.}~\bibnamefont
  {Dupont}}, \bibinfo {author} {\bibfnamefont {P.~B.}\ \bibnamefont {Corkum}},
  \bibinfo {author} {\bibfnamefont {H.~C.}\ \bibnamefont {Liu}}, \bibinfo
  {author} {\bibfnamefont {M.}~\bibnamefont {Buchanan}}, \ and\ \bibinfo
  {author} {\bibfnamefont {Z.~R.}\ \bibnamefont {Wasilewski}},\ }\href@noop {}
  {\bibfield  {journal} {\bibinfo  {journal} {Physical Review Letters}\
  }\textbf {\bibinfo {volume} {74}},\ \bibinfo {pages} {3596} (\bibinfo {year}
  {1995})}\BibitemShut {NoStop}%
\bibitem [{\citenamefont {Atanasov}\ \emph {et~al.}(1996)\citenamefont
  {Atanasov}, \citenamefont {Hache}, \citenamefont {Hughes}, \citenamefont {van
  Driel},\ and\ \citenamefont {Sipe}}]{Atanasov96}%
  \BibitemOpen
  \bibfield  {author} {\bibinfo {author} {\bibfnamefont {R.}~\bibnamefont
  {Atanasov}}, \bibinfo {author} {\bibfnamefont {A.}~\bibnamefont {Hache}},
  \bibinfo {author} {\bibfnamefont {J.~L.~P.}\ \bibnamefont {Hughes}}, \bibinfo
  {author} {\bibfnamefont {H.~M.}\ \bibnamefont {van Driel}}, \ and\ \bibinfo
  {author} {\bibfnamefont {J.~E.}\ \bibnamefont {Sipe}},\ }\href@noop {}
  {\bibfield  {journal} {\bibinfo  {journal} {Phys. Rev. Lett.}\ }\textbf
  {\bibinfo {volume} {76}},\ \bibinfo {pages} {1703} (\bibinfo {year}
  {1996})}\BibitemShut {NoStop}%
\bibitem [{\citenamefont {Hache}\ \emph {et~al.}(1997)\citenamefont {Hache},
  \citenamefont {Kostoulas}, \citenamefont {Atanasov}, \citenamefont {Hughes},\
  and\ \citenamefont {Sipe}}]{Hache97}%
  \BibitemOpen
  \bibfield  {author} {\bibinfo {author} {\bibfnamefont {A.}~\bibnamefont
  {Hache}}, \bibinfo {author} {\bibfnamefont {Y.}~\bibnamefont {Kostoulas}},
  \bibinfo {author} {\bibfnamefont {R.}~\bibnamefont {Atanasov}}, \bibinfo
  {author} {\bibfnamefont {J.~J.~P.}\ \bibnamefont {Hughes}}, \ and\ \bibinfo
  {author} {\bibfnamefont {J.~E.}\ \bibnamefont {Sipe}},\ }\href@noop {}
  {\bibfield  {journal} {\bibinfo  {journal} {Phys. Rev. Lett.}\ }\textbf
  {\bibinfo {volume} {78}},\ \bibinfo {pages} {306} (\bibinfo {year}
  {1997})}\BibitemShut {NoStop}%
\bibitem [{\citenamefont {Rioux}\ and\ \citenamefont {Sipe}(2012)}]{Rioux12}%
  \BibitemOpen
  \bibfield  {author} {\bibinfo {author} {\bibfnamefont {J.}~\bibnamefont
  {Rioux}}\ and\ \bibinfo {author} {\bibfnamefont {J.~E.}\ \bibnamefont
  {Sipe}},\ }\href@noop {} {\bibfield  {journal} {\bibinfo  {journal} {Physica
  E}\ }\textbf {\bibinfo {volume} {45}},\ \bibinfo {pages} {1} (\bibinfo {year}
  {2012})}\BibitemShut {NoStop}%
\bibitem [{\citenamefont {Sun}\ \emph {et~al.}(2010)\citenamefont {Sun},
  \citenamefont {Divin}, \citenamefont {Rioux}, \citenamefont {Sipe},
  \citenamefont {Berger}, \citenamefont {de~Heer}, \citenamefont {First},\ and\
  \citenamefont {Norris}}]{Sun10}%
  \BibitemOpen
  \bibfield  {author} {\bibinfo {author} {\bibfnamefont {D.}~\bibnamefont
  {Sun}}, \bibinfo {author} {\bibfnamefont {C.}~\bibnamefont {Divin}}, \bibinfo
  {author} {\bibfnamefont {J.}~\bibnamefont {Rioux}}, \bibinfo {author}
  {\bibfnamefont {J.~E.}\ \bibnamefont {Sipe}}, \bibinfo {author}
  {\bibfnamefont {C.}~\bibnamefont {Berger}}, \bibinfo {author} {\bibfnamefont
  {W.~A.}\ \bibnamefont {de~Heer}}, \bibinfo {author} {\bibfnamefont {P.~N.}\
  \bibnamefont {First}}, \ and\ \bibinfo {author} {\bibfnamefont {T.~B.}\
  \bibnamefont {Norris}},\ }\href@noop {} {\bibfield  {journal} {\bibinfo
  {journal} {Nano Lett.}\ }\textbf {\bibinfo {volume} {10}},\ \bibinfo {pages}
  {1293} (\bibinfo {year} {2010})}\BibitemShut {NoStop}%
\bibitem [{\citenamefont {Rioux}\ \emph {et~al.}(2011)\citenamefont {Rioux},
  \citenamefont {Burkard},\ and\ \citenamefont {Sipe}}]{Rioux11}%
  \BibitemOpen
  \bibfield  {author} {\bibinfo {author} {\bibfnamefont {J.}~\bibnamefont
  {Rioux}}, \bibinfo {author} {\bibfnamefont {G.}~\bibnamefont {Burkard}}, \
  and\ \bibinfo {author} {\bibfnamefont {J.~E.}\ \bibnamefont {Sipe}},\
  }\href@noop {} {\bibfield  {journal} {\bibinfo  {journal} {Phys. Rev. B}\
  }\textbf {\bibinfo {volume} {83}},\ \bibinfo {pages} {195406} (\bibinfo
  {year} {2011})}\BibitemShut {NoStop}%
\bibitem [{\citenamefont {Rao}\ and\ \citenamefont {Sipe}(2012)}]{Rao12}%
  \BibitemOpen
  \bibfield  {author} {\bibinfo {author} {\bibfnamefont {K.~M.}\ \bibnamefont
  {Rao}}\ and\ \bibinfo {author} {\bibfnamefont {J.~E.}\ \bibnamefont {Sipe}},\
  }\href@noop {} {\bibfield  {journal} {\bibinfo  {journal} {Phys. Rev. B}\
  }\textbf {\bibinfo {volume} {86}},\ \bibinfo {pages} {115427} (\bibinfo
  {year} {2012})}\BibitemShut {NoStop}%
\bibitem [{\citenamefont {Muniz}\ and\ \citenamefont {Sipe}(2014)}]{Muniz14}%
  \BibitemOpen
  \bibfield  {author} {\bibinfo {author} {\bibfnamefont {R.~A.}\ \bibnamefont
  {Muniz}}\ and\ \bibinfo {author} {\bibfnamefont {J.~E.}\ \bibnamefont
  {Sipe}},\ }\href@noop {} {\bibfield  {journal} {\bibinfo  {journal} {Phys.
  Rev. B}\ }\textbf {\bibinfo {volume} {89}},\ \bibinfo {pages} {205113}
  (\bibinfo {year} {2014})}\BibitemShut {NoStop}%
\bibitem [{\citenamefont {Bas}\ \emph {et~al.}(2015)\citenamefont {Bas},
  \citenamefont {Vargas-Velez}, \citenamefont {Babakiray}, \citenamefont
  {Johnson}, \citenamefont {Borisov}, \citenamefont {Stanescu}, \citenamefont
  {Lederman},\ and\ \citenamefont {Bristow}}]{Bas15}%
  \BibitemOpen
  \bibfield  {author} {\bibinfo {author} {\bibfnamefont {D.~A.}\ \bibnamefont
  {Bas}}, \bibinfo {author} {\bibfnamefont {K.}~\bibnamefont {Vargas-Velez}},
  \bibinfo {author} {\bibfnamefont {S.}~\bibnamefont {Babakiray}}, \bibinfo
  {author} {\bibfnamefont {T.~A.}\ \bibnamefont {Johnson}}, \bibinfo {author}
  {\bibfnamefont {P.}~\bibnamefont {Borisov}}, \bibinfo {author} {\bibfnamefont
  {T.~D.}\ \bibnamefont {Stanescu}}, \bibinfo {author} {\bibfnamefont
  {D.}~\bibnamefont {Lederman}}, \ and\ \bibinfo {author} {\bibfnamefont
  {A.~D.}\ \bibnamefont {Bristow}},\ }\href@noop {} {\bibfield  {journal}
  {\bibinfo  {journal} {Appl. Phys. Lett.}\ }\textbf {\bibinfo {volume}
  {106}},\ \bibinfo {pages} {041109} (\bibinfo {year} {2015})}\BibitemShut
  {NoStop}%
\bibitem [{\citenamefont {Bas}\ \emph {et~al.}(2016)\citenamefont {Bas},
  \citenamefont {Muniz}, \citenamefont {Babakiray}, \citenamefont {Lederman},
  \citenamefont {Sipe},\ and\ \citenamefont {Bristow}}]{Bas16}%
  \BibitemOpen
  \bibfield  {author} {\bibinfo {author} {\bibfnamefont {D.~A.}\ \bibnamefont
  {Bas}}, \bibinfo {author} {\bibfnamefont {R.~A.}\ \bibnamefont {Muniz}},
  \bibinfo {author} {\bibfnamefont {S.}~\bibnamefont {Babakiray}}, \bibinfo
  {author} {\bibfnamefont {D.}~\bibnamefont {Lederman}}, \bibinfo {author}
  {\bibfnamefont {J.~E.}\ \bibnamefont {Sipe}}, \ and\ \bibinfo {author}
  {\bibfnamefont {A.~D.}\ \bibnamefont {Bristow}},\ }\href@noop {} {\bibfield
  {journal} {\bibinfo  {journal} {Opt. Express}\ }\textbf {\bibinfo {volume}
  {24}},\ \bibinfo {pages} {23583} (\bibinfo {year} {2016})}\BibitemShut
  {NoStop}%
\bibitem [{\citenamefont {Muniz}\ and\ \citenamefont {Sipe}(2015)}]{Muniz15}%
  \BibitemOpen
  \bibfield  {author} {\bibinfo {author} {\bibfnamefont {R.~A.}\ \bibnamefont
  {Muniz}}\ and\ \bibinfo {author} {\bibfnamefont {J.~E.}\ \bibnamefont
  {Sipe}},\ }\href@noop {} {\bibfield  {journal} {\bibinfo  {journal} {Phys.
  Rev. B}\ }\textbf {\bibinfo {volume} {91}},\ \bibinfo {pages} {085404}
  (\bibinfo {year} {2015})}\BibitemShut {NoStop}%
\bibitem [{\citenamefont {Cui}\ and\ \citenamefont {Zhao}(2015)}]{Cui15}%
  \BibitemOpen
  \bibfield  {author} {\bibinfo {author} {\bibfnamefont {Q.}~\bibnamefont
  {Cui}}\ and\ \bibinfo {author} {\bibfnamefont {H.}~\bibnamefont {Zhao}},\
  }\href@noop {} {\bibfield  {journal} {\bibinfo  {journal} {ACS Nano}\
  }\textbf {\bibinfo {volume} {9}},\ \bibinfo {pages} {3935} (\bibinfo {year}
  {2015})}\BibitemShut {NoStop}%
\bibitem [{\citenamefont {Mahon}\ \emph {et~al.}(2018)\citenamefont {Mahon},
  \citenamefont {Muniz},\ and\ \citenamefont {Sipe}}]{Mahon18}%
  \BibitemOpen
  \bibfield  {author} {\bibinfo {author} {\bibfnamefont {P.~T.}\ \bibnamefont
  {Mahon}}, \bibinfo {author} {\bibfnamefont {R.~A.}\ \bibnamefont {Muniz}}, \
  and\ \bibinfo {author} {\bibfnamefont {J.~E.}\ \bibnamefont {Sipe}},\
  }\href@noop {} {\bibfield  {journal} {\bibinfo  {journal} {arXiv:1810.09971
  [cond-mat.mes-hall]}\ } (\bibinfo {year} {2018})}\BibitemShut {NoStop}%
\bibitem [{\citenamefont {Bhat}\ and\ \citenamefont {Sipe}(2000)}]{Bhat00}%
  \BibitemOpen
  \bibfield  {author} {\bibinfo {author} {\bibfnamefont {R.~D.~R.}\
  \bibnamefont {Bhat}}\ and\ \bibinfo {author} {\bibfnamefont {J.~E.}\
  \bibnamefont {Sipe}},\ }\href@noop {} {\bibfield  {journal} {\bibinfo
  {journal} {Phys. Rev. Lett.}\ }\textbf {\bibinfo {volume} {85}},\ \bibinfo
  {pages} {5432} (\bibinfo {year} {2000})}\BibitemShut {NoStop}%
\bibitem [{\citenamefont {Stevens}\ \emph {et~al.}(2002)\citenamefont
  {Stevens}, \citenamefont {Smirl}, \citenamefont {Bhat}, \citenamefont
  {Sipe},\ and\ \citenamefont {van Driel}}]{Stevens02}%
  \BibitemOpen
  \bibfield  {author} {\bibinfo {author} {\bibfnamefont {M.~J.}\ \bibnamefont
  {Stevens}}, \bibinfo {author} {\bibfnamefont {A.~L.}\ \bibnamefont {Smirl}},
  \bibinfo {author} {\bibfnamefont {R.~D.~R.}\ \bibnamefont {Bhat}}, \bibinfo
  {author} {\bibfnamefont {J.~E.}\ \bibnamefont {Sipe}}, \ and\ \bibinfo
  {author} {\bibfnamefont {H.~M.}\ \bibnamefont {van Driel}},\ }\href@noop {}
  {\bibfield  {journal} {\bibinfo  {journal} {J. Appl. Phys.}\ }\textbf
  {\bibinfo {volume} {91}},\ \bibinfo {pages} {4382} (\bibinfo {year}
  {2002})}\BibitemShut {NoStop}%
\bibitem [{\citenamefont {Stevens}\ \emph {et~al.}(2003)\citenamefont
  {Stevens}, \citenamefont {Smirl}, \citenamefont {Bhat}, \citenamefont
  {Najmaie}, \citenamefont {Sipe},\ and\ \citenamefont {van
  Driel}}]{Stevens03}%
  \BibitemOpen
  \bibfield  {author} {\bibinfo {author} {\bibfnamefont {M.~J.}\ \bibnamefont
  {Stevens}}, \bibinfo {author} {\bibfnamefont {A.~L.}\ \bibnamefont {Smirl}},
  \bibinfo {author} {\bibfnamefont {R.~D.~R.}\ \bibnamefont {Bhat}}, \bibinfo
  {author} {\bibfnamefont {A.}~\bibnamefont {Najmaie}}, \bibinfo {author}
  {\bibfnamefont {J.~E.}\ \bibnamefont {Sipe}}, \ and\ \bibinfo {author}
  {\bibfnamefont {H.~M.}\ \bibnamefont {van Driel}},\ }\href@noop {} {\bibfield
   {journal} {\bibinfo  {journal} {Phys. Rev. Lett.}\ }\textbf {\bibinfo
  {volume} {90}},\ \bibinfo {pages} {136603} (\bibinfo {year}
  {2003})}\BibitemShut {NoStop}%
\bibitem [{\citenamefont {Hubner}\ \emph {et~al.}(2003)\citenamefont {Hubner},
  \citenamefont {Ruhle}, \citenamefont {Klude}, \citenamefont {Hommel},
  \citenamefont {Bhat}, \citenamefont {Sipe},\ and\ \citenamefont {van
  Driel}}]{Hubner03}%
  \BibitemOpen
  \bibfield  {author} {\bibinfo {author} {\bibfnamefont {J.}~\bibnamefont
  {Hubner}}, \bibinfo {author} {\bibfnamefont {W.~W.}\ \bibnamefont {Ruhle}},
  \bibinfo {author} {\bibfnamefont {M.}~\bibnamefont {Klude}}, \bibinfo
  {author} {\bibfnamefont {D.}~\bibnamefont {Hommel}}, \bibinfo {author}
  {\bibfnamefont {R.~D.~R.}\ \bibnamefont {Bhat}}, \bibinfo {author}
  {\bibfnamefont {J.~E.}\ \bibnamefont {Sipe}}, \ and\ \bibinfo {author}
  {\bibfnamefont {H.~M.}\ \bibnamefont {van Driel}},\ }\href@noop {} {\bibfield
   {journal} {\bibinfo  {journal} {Phys. Rev. Lett.}\ }\textbf {\bibinfo
  {volume} {90}},\ \bibinfo {pages} {216601} (\bibinfo {year}
  {2003})}\BibitemShut {NoStop}%
\bibitem [{\citenamefont {Zhao}\ \emph {et~al.}(2006)\citenamefont {Zhao},
  \citenamefont {Loren}, \citenamefont {van Driel},\ and\ \citenamefont
  {Smirl}}]{Zhao06}%
  \BibitemOpen
  \bibfield  {author} {\bibinfo {author} {\bibfnamefont {H.}~\bibnamefont
  {Zhao}}, \bibinfo {author} {\bibfnamefont {E.~J.}\ \bibnamefont {Loren}},
  \bibinfo {author} {\bibfnamefont {H.~M.}\ \bibnamefont {van Driel}}, \ and\
  \bibinfo {author} {\bibfnamefont {A.~L.}\ \bibnamefont {Smirl}},\ }\href@noop
  {} {\bibfield  {journal} {\bibinfo  {journal} {Phys. Rev. Lett.}\ }\textbf
  {\bibinfo {volume} {96}},\ \bibinfo {pages} {246601} (\bibinfo {year}
  {2006})}\BibitemShut {NoStop}%
\bibitem [{\citenamefont {Salazar}\ \emph {et~al.}(2016)\citenamefont
  {Salazar}, \citenamefont {Cheng},\ and\ \citenamefont {Sipe}}]{Salazar16}%
  \BibitemOpen
  \bibfield  {author} {\bibinfo {author} {\bibfnamefont {C.}~\bibnamefont
  {Salazar}}, \bibinfo {author} {\bibfnamefont {J.~L.}\ \bibnamefont {Cheng}},
  \ and\ \bibinfo {author} {\bibfnamefont {J.~E.}\ \bibnamefont {Sipe}},\
  }\href@noop {} {\bibfield  {journal} {\bibinfo  {journal} {Phys. Rev. B}\
  }\textbf {\bibinfo {volume} {93}},\ \bibinfo {pages} {075442} (\bibinfo
  {year} {2016})}\BibitemShut {NoStop}%
\bibitem [{\citenamefont {Behnia}(2012)}]{Behnia12}%
  \BibitemOpen
  \bibfield  {author} {\bibinfo {author} {\bibfnamefont {K.}~\bibnamefont
  {Behnia}},\ }\href@noop {} {\bibfield  {journal} {\bibinfo  {journal} {Nature
  Nanotechnology}\ }\textbf {\bibinfo {volume} {7}},\ \bibinfo {pages} {488}
  (\bibinfo {year} {2012})}\BibitemShut {NoStop}%
\bibitem [{\citenamefont {Roos}\ \emph
  {et~al.}(2005{\natexlab{a}})\citenamefont {Roos}, \citenamefont {Li},
  \citenamefont {Smith}, \citenamefont {Pipis}, \citenamefont {Fortier},\ and\
  \citenamefont {Cundiff}}]{Roos05optlet}%
  \BibitemOpen
  \bibfield  {author} {\bibinfo {author} {\bibfnamefont {P.~A.}\ \bibnamefont
  {Roos}}, \bibinfo {author} {\bibfnamefont {X.}~\bibnamefont {Li}}, \bibinfo
  {author} {\bibfnamefont {R.~P.}\ \bibnamefont {Smith}}, \bibinfo {author}
  {\bibfnamefont {J.~A.}\ \bibnamefont {Pipis}}, \bibinfo {author}
  {\bibfnamefont {T.~M.}\ \bibnamefont {Fortier}}, \ and\ \bibinfo {author}
  {\bibfnamefont {S.~T.}\ \bibnamefont {Cundiff}},\ }\href@noop {} {\bibfield
  {journal} {\bibinfo  {journal} {Opt. Lett.}\ }\textbf {\bibinfo {volume}
  {30}},\ \bibinfo {pages} {735} (\bibinfo {year}
  {2005}{\natexlab{a}})}\BibitemShut {NoStop}%
\bibitem [{\citenamefont {Roos}\ and\ \citenamefont
  {Cundiff}(2005)}]{Roos05laser}%
  \BibitemOpen
  \bibfield  {author} {\bibinfo {author} {\bibfnamefont {P.~A.}\ \bibnamefont
  {Roos}}\ and\ \bibinfo {author} {\bibfnamefont {S.~T.}\ \bibnamefont
  {Cundiff}},\ }\href@noop {} {\bibfield  {journal} {\bibinfo  {journal} {Laser
  Physics}\ }\textbf {\bibinfo {volume} {15}},\ \bibinfo {pages} {769}
  (\bibinfo {year} {2005})}\BibitemShut {NoStop}%
\bibitem [{\citenamefont {Smith}\ \emph {et~al.}(2007)\citenamefont {Smith},
  \citenamefont {Roos}, \citenamefont {Wahlstrand}, \citenamefont {Pipis},
  \citenamefont {Rivas},\ and\ \citenamefont {Cundiff}}]{Smith07}%
  \BibitemOpen
  \bibfield  {author} {\bibinfo {author} {\bibfnamefont {R.}~\bibnamefont
  {Smith}}, \bibinfo {author} {\bibfnamefont {P.~A.}\ \bibnamefont {Roos}},
  \bibinfo {author} {\bibfnamefont {J.~K.}\ \bibnamefont {Wahlstrand}},
  \bibinfo {author} {\bibfnamefont {J.~A.}\ \bibnamefont {Pipis}}, \bibinfo
  {author} {\bibfnamefont {M.~B.}\ \bibnamefont {Rivas}}, \ and\ \bibinfo
  {author} {\bibfnamefont {S.~T.}\ \bibnamefont {Cundiff}},\ }\href@noop {}
  {\bibfield  {journal} {\bibinfo  {journal} {J. Res. Nat. Inst. Sci. Tech.}\
  }\textbf {\bibinfo {volume} {112}},\ \bibinfo {pages} {289} (\bibinfo {year}
  {2007})}\BibitemShut {NoStop}%
\bibitem [{\citenamefont {Roos}\ \emph {et~al.}(2004)\citenamefont {Roos},
  \citenamefont {Li}, \citenamefont {Pipis},\ and\ \citenamefont
  {Cundiff}}]{Roos04}%
  \BibitemOpen
  \bibfield  {author} {\bibinfo {author} {\bibfnamefont {P.~A.}\ \bibnamefont
  {Roos}}, \bibinfo {author} {\bibfnamefont {X.}~\bibnamefont {Li}}, \bibinfo
  {author} {\bibfnamefont {J.~R.}\ \bibnamefont {Pipis}}, \ and\ \bibinfo
  {author} {\bibfnamefont {S.~T.}\ \bibnamefont {Cundiff}},\ }\href@noop {}
  {\bibfield  {journal} {\bibinfo  {journal} {Opt. Exp.}\ }\textbf {\bibinfo
  {volume} {12}},\ \bibinfo {pages} {4225} (\bibinfo {year}
  {2004})}\BibitemShut {NoStop}%
\bibitem [{\citenamefont {Fortier}\ \emph {et~al.}(2004)\citenamefont
  {Fortier}, \citenamefont {Roos}, \citenamefont {Jones}, \citenamefont
  {Cundiff}, \citenamefont {Bhat},\ and\ \citenamefont {Sipe}}]{Fortier04}%
  \BibitemOpen
  \bibfield  {author} {\bibinfo {author} {\bibfnamefont {T.~M.}\ \bibnamefont
  {Fortier}}, \bibinfo {author} {\bibfnamefont {P.~A.}\ \bibnamefont {Roos}},
  \bibinfo {author} {\bibfnamefont {D.~J.}\ \bibnamefont {Jones}}, \bibinfo
  {author} {\bibfnamefont {S.~T.}\ \bibnamefont {Cundiff}}, \bibinfo {author}
  {\bibfnamefont {R.~D.~R.}\ \bibnamefont {Bhat}}, \ and\ \bibinfo {author}
  {\bibfnamefont {J.~E.}\ \bibnamefont {Sipe}},\ }\href@noop {} {\bibfield
  {journal} {\bibinfo  {journal} {Phys. Rev. Lett.}\ }\textbf {\bibinfo
  {volume} {92}},\ \bibinfo {pages} {147403} (\bibinfo {year}
  {2004})}\BibitemShut {NoStop}%
\bibitem [{\citenamefont {Cundiff}\ and\ \citenamefont {Ye}(2003)}]{Cundiff03}%
  \BibitemOpen
  \bibfield  {author} {\bibinfo {author} {\bibfnamefont {S.~T.}\ \bibnamefont
  {Cundiff}}\ and\ \bibinfo {author} {\bibfnamefont {J.}~\bibnamefont {Ye}},\
  }\href@noop {} {\bibfield  {journal} {\bibinfo  {journal} {Rev. Mod. Phys.}\
  }\textbf {\bibinfo {volume} {75}},\ \bibinfo {pages} {325} (\bibinfo {year}
  {2003})}\BibitemShut {NoStop}%
\bibitem [{\citenamefont {Ye}\ and\ \citenamefont {Cundiff}(2005)}]{Ye05}%
  \BibitemOpen
  \bibfield  {author} {\bibinfo {author} {\bibfnamefont {J.}~\bibnamefont
  {Ye}}\ and\ \bibinfo {author} {\bibfnamefont {S.}~\bibnamefont {Cundiff}},\
  }\href@noop {} {\emph {\bibinfo {title} {Femtosecond Optical Frequency Comb:
  Principle, Operation, and Applications}}}\ (\bibinfo  {publisher} {Springer,
  Norwell, MA},\ \bibinfo {year} {2005})\BibitemShut {NoStop}%
\bibitem [{\citenamefont {Roos}\ \emph
  {et~al.}(2005{\natexlab{b}})\citenamefont {Roos}, \citenamefont {Li},
  \citenamefont {Pipis}, \citenamefont {Fortier}, \citenamefont {Cundiff},
  \citenamefont {Bhat},\ and\ \citenamefont {Sipe}}]{Roos05josaB}%
  \BibitemOpen
  \bibfield  {author} {\bibinfo {author} {\bibfnamefont {P.~A.}\ \bibnamefont
  {Roos}}, \bibinfo {author} {\bibfnamefont {X.}~\bibnamefont {Li}}, \bibinfo
  {author} {\bibfnamefont {J.~A.}\ \bibnamefont {Pipis}}, \bibinfo {author}
  {\bibfnamefont {T.~M.}\ \bibnamefont {Fortier}}, \bibinfo {author}
  {\bibfnamefont {S.~T.}\ \bibnamefont {Cundiff}}, \bibinfo {author}
  {\bibfnamefont {R.~D.~R.}\ \bibnamefont {Bhat}}, \ and\ \bibinfo {author}
  {\bibfnamefont {J.~E.}\ \bibnamefont {Sipe}},\ }\href@noop {} {\bibfield
  {journal} {\bibinfo  {journal} {J. Opt. Soc. Am. B}\ }\textbf {\bibinfo
  {volume} {22}},\ \bibinfo {pages} {362} (\bibinfo {year}
  {2005}{\natexlab{b}})}\BibitemShut {NoStop}%
\bibitem [{\citenamefont {Nastos}\ \emph {et~al.}(2007)\citenamefont {Nastos},
  \citenamefont {Rioux}, \citenamefont {Strimas-Mackey}, \citenamefont
  {Mendoza},\ and\ \citenamefont {Sipe}}]{Nastos07}%
  \BibitemOpen
  \bibfield  {author} {\bibinfo {author} {\bibfnamefont {F.}~\bibnamefont
  {Nastos}}, \bibinfo {author} {\bibfnamefont {J.}~\bibnamefont {Rioux}},
  \bibinfo {author} {\bibfnamefont {M.}~\bibnamefont {Strimas-Mackey}},
  \bibinfo {author} {\bibfnamefont {B.~S.}\ \bibnamefont {Mendoza}}, \ and\
  \bibinfo {author} {\bibfnamefont {J.~E.}\ \bibnamefont {Sipe}},\ }\href@noop
  {} {\bibfield  {journal} {\bibinfo  {journal} {Phys. Rev. B}\ }\textbf
  {\bibinfo {volume} {76}},\ \bibinfo {pages} {205113} (\bibinfo {year}
  {2007})}\BibitemShut {NoStop}%
\bibitem [{\citenamefont {Sternemann}\ \emph {et~al.}(2013)\citenamefont
  {Sternemann}, \citenamefont {Jostmeier}, \citenamefont {Ruppert},
  \citenamefont {Duc}, \citenamefont {Meier},\ and\ \citenamefont
  {Betz}}]{Sternemann13}%
  \BibitemOpen
  \bibfield  {author} {\bibinfo {author} {\bibfnamefont {E.}~\bibnamefont
  {Sternemann}}, \bibinfo {author} {\bibfnamefont {T.}~\bibnamefont
  {Jostmeier}}, \bibinfo {author} {\bibfnamefont {C.}~\bibnamefont {Ruppert}},
  \bibinfo {author} {\bibfnamefont {H.~T.}\ \bibnamefont {Duc}}, \bibinfo
  {author} {\bibfnamefont {T.}~\bibnamefont {Meier}}, \ and\ \bibinfo {author}
  {\bibfnamefont {M.}~\bibnamefont {Betz}},\ }\href@noop {} {\bibfield
  {journal} {\bibinfo  {journal} {Phys. Rev. B}\ }\textbf {\bibinfo {volume}
  {88}},\ \bibinfo {pages} {165204} (\bibinfo {year} {2013})}\BibitemShut
  {NoStop}%
\bibitem [{\citenamefont {Sternemann}\ \emph {et~al.}(2016)\citenamefont
  {Sternemann}, \citenamefont {Jostmeier}, \citenamefont {Ruppert},
  \citenamefont {Thunich}, \citenamefont {Duc}, \citenamefont {Podzimski},
  \citenamefont {Meier},\ and\ \citenamefont {Betz}}]{Sternemann16}%
  \BibitemOpen
  \bibfield  {author} {\bibinfo {author} {\bibfnamefont {E.}~\bibnamefont
  {Sternemann}}, \bibinfo {author} {\bibfnamefont {T.}~\bibnamefont
  {Jostmeier}}, \bibinfo {author} {\bibfnamefont {C.}~\bibnamefont {Ruppert}},
  \bibinfo {author} {\bibfnamefont {S.}~\bibnamefont {Thunich}}, \bibinfo
  {author} {\bibfnamefont {H.~T.}\ \bibnamefont {Duc}}, \bibinfo {author}
  {\bibfnamefont {R.}~\bibnamefont {Podzimski}}, \bibinfo {author}
  {\bibfnamefont {T.}~\bibnamefont {Meier}}, \ and\ \bibinfo {author}
  {\bibfnamefont {M.}~\bibnamefont {Betz}},\ }\href@noop {} {\bibfield
  {journal} {\bibinfo  {journal} {Applied Physics B}\ }\textbf {\bibinfo
  {volume} {122}},\ \bibinfo {pages} {1} (\bibinfo {year} {2016})}\BibitemShut
  {NoStop}%
\bibitem [{\citenamefont {Wang}\ \emph {et~al.}(2018)\citenamefont {Wang},
  \citenamefont {Muniz}, \citenamefont {Sipe},\ and\ \citenamefont
  {Cundiff}}]{Wang17}%
  \BibitemOpen
  \bibfield  {author} {\bibinfo {author} {\bibfnamefont {K.}~\bibnamefont
  {Wang}}, \bibinfo {author} {\bibfnamefont {R.~A.}\ \bibnamefont {Muniz}},
  \bibinfo {author} {\bibfnamefont {J.~E.}\ \bibnamefont {Sipe}}, \ and\
  \bibinfo {author} {\bibfnamefont {S.~T.}\ \bibnamefont {Cundiff}},\
  }\href@noop {} {\bibfield  {journal} {\bibinfo  {journal} {arXiv:1808.07523
  [cond-mat.mes-hall]}\ } (\bibinfo {year} {2018})}\BibitemShut {NoStop}%
\bibitem [{Note1()}]{Note1}%
  \BibitemOpen
  \bibinfo {note} {The denominators in Eqs.~\protect \textup {\hbox
  {\mathsurround \z@ \protect \normalfont (\ignorespaces \ref
  {eq:R2coefSimp}\unskip \@@italiccorr )}} and \protect \textup {\hbox
  {\mathsurround \z@ \protect \normalfont (\ignorespaces \ref
  {eq:R3coefSimp}\unskip \@@italiccorr )}} do not lead to any divergences
  because of the assumption that $2\hbar \omega $ is below the
  gap.}\BibitemShut {Stop}%
\bibitem [{\citenamefont {Richard}\ \emph {et~al.}(2004)\citenamefont
  {Richard}, \citenamefont {Aniel},\ and\ \citenamefont {Fishman}}]{richard04}%
  \BibitemOpen
  \bibfield  {author} {\bibinfo {author} {\bibfnamefont {S.}~\bibnamefont
  {Richard}}, \bibinfo {author} {\bibfnamefont {F.}~\bibnamefont {Aniel}}, \
  and\ \bibinfo {author} {\bibfnamefont {G.}~\bibnamefont {Fishman}},\
  }\href@noop {} {\bibfield  {journal} {\bibinfo  {journal} {Phys. Rev. B}\
  }\textbf {\bibinfo {volume} {70}},\ \bibinfo {pages} {235204} (\bibinfo
  {year} {2004})}\BibitemShut {NoStop}%
\bibitem [{\citenamefont {Salazar}\ \emph {et~al.}(2017)\citenamefont
  {Salazar}, \citenamefont {Muniz},\ and\ \citenamefont {Sipe}}]{Salazar17}%
  \BibitemOpen
  \bibfield  {author} {\bibinfo {author} {\bibfnamefont {C.}~\bibnamefont
  {Salazar}}, \bibinfo {author} {\bibfnamefont {R.~A.}\ \bibnamefont {Muniz}},
  \ and\ \bibinfo {author} {\bibfnamefont {J.~E.}\ \bibnamefont {Sipe}},\
  }\href@noop {} {\bibfield  {journal} {\bibinfo  {journal} {Phys. Rev.
  Materials}\ }\textbf {\bibinfo {volume} {1}},\ \bibinfo {pages} {054006}
  (\bibinfo {year} {2017})}\BibitemShut {NoStop}%
\bibitem [{\citenamefont {Fraj}\ \emph {et~al.}(2007)\citenamefont {Fraj},
  \citenamefont {Saidi}, \citenamefont {Radhia},\ and\ \citenamefont
  {Boujdaria}}]{fraj07}%
  \BibitemOpen
  \bibfield  {author} {\bibinfo {author} {\bibfnamefont {N.}~\bibnamefont
  {Fraj}}, \bibinfo {author} {\bibfnamefont {I.}~\bibnamefont {Saidi}},
  \bibinfo {author} {\bibfnamefont {S.~B.}\ \bibnamefont {Radhia}}, \ and\
  \bibinfo {author} {\bibfnamefont {K.}~\bibnamefont {Boujdaria}},\ }\href@noop
  {} {\bibfield  {journal} {\bibinfo  {journal} {J. Appl. Phys.}\ }\textbf
  {\bibinfo {volume} {102}},\ \bibinfo {pages} {053703} (\bibinfo {year}
  {2007})}\BibitemShut {NoStop}%
\bibitem [{Note2()}]{Note2}%
  \BibitemOpen
  \bibinfo {note} {The corrections that are quadratic on the stoichiometry
  parameter $\alpha $ are small, and do not lead to significant changes in the
  band energies within a tolerance given by room temperature.}\BibitemShut
  {Stop}%
\bibitem [{\citenamefont {Rioux}(2011)}]{Rioux11th}%
  \BibitemOpen
  \bibfield  {author} {\bibinfo {author} {\bibfnamefont {J.}~\bibnamefont
  {Rioux}},\ }\emph {\bibinfo {title} {Full-band structure calculations of
  optical injection in semiconductors: Investigations of one-color, two-color,
  and pump-probe scenarios}},\ \href@noop {} {Ph.D. thesis},\ \bibinfo
  {school} {University of Toronto} (\bibinfo {year} {2011})\BibitemShut
  {NoStop}%
\bibitem [{Note3()}]{Note3}%
  \BibitemOpen
  \bibinfo {note} {If the band gap of the material is smaller than $5\omega /2$
  there would be an additional contribution to $n_{2}$ proportional to
  $E_{-\omega }E_{-3\omega /2}E_{\omega }E_{3\omega /2}$, but these carriers do
  not contribute to the interference between 2PA and 3PA, which is our main
  interest.}\BibitemShut {Stop}%
\bibitem [{Note4()}]{Note4}%
  \BibitemOpen
  \bibinfo {note} {Notice that assuming constant injection rates, which is a
  reasonable assumption at least during some time interval, the definition of
  the swarm velocity is equivalent to $ {\protect \bf v}_{\protect \rm swarm}=
  <{\protect \bf J}> /e <n> $.}\BibitemShut {Stop}%
\bibitem [{\citenamefont {Bhat}\ and\ \citenamefont {Sipe}(2005)}]{bhat05}%
  \BibitemOpen
  \bibfield  {author} {\bibinfo {author} {\bibfnamefont {R.~D.~R.}\
  \bibnamefont {Bhat}}\ and\ \bibinfo {author} {\bibfnamefont {J.~E.}\
  \bibnamefont {Sipe}},\ }\href@noop {} {\bibfield  {journal} {\bibinfo
  {journal} {Phys. Rev. B}\ }\textbf {\bibinfo {volume} {72}},\ \bibinfo
  {pages} {075205} (\bibinfo {year} {2005})}\BibitemShut {NoStop}%
\bibitem [{\citenamefont {Sauter}(1996)}]{Sauter96}%
  \BibitemOpen
  \bibfield  {author} {\bibinfo {author} {\bibfnamefont {E.~G.}\ \bibnamefont
  {Sauter}},\ }\href@noop {} {\emph {\bibinfo {title} {Nonlinear Optics}}}\
  (\bibinfo  {publisher} {Wiley-Interscience, New York},\ \bibinfo {year}
  {1996})\BibitemShut {NoStop}%
\end{thebibliography}%

\end{document}